\documentclass[11pt]{article}
\usepackage{jcappub}
\usepackage{bm}
\usepackage{color}
\usepackage{graphicx}
\usepackage{multirow}
\usepackage{epstopdf}
\usepackage{geometry}
\usepackage{bbold}
\usepackage{hyperref}
\usepackage{fancyvrb}

\usepackage{bm}

\newcommand{\qmsq}{\qmsq}
\renewcommand{\qmsq}{\q^2}

\newcommand{\be}{\begin{equation}}
\newcommand{\ee}{\end{equation}}
\newcommand{\een}{\end{subequations}}
\newcommand{\ben}{\begin{subequations}}
\newcommand{\beq}{\begin{eqalignno}}
\newcommand{\eeq}{\end{eqalignno}}

\newcommand{\lsim}{\mathrel{\mathop{\kern 0pt \rlap
      {\raise.2ex\hbox{$<$}}}\lower.9ex\hbox{\kern-.190em $ \sim$}}}
\newcommand{\gsim}{\mathrel{\mathop{\kern 0pt
      \rlap{\raise.2ex\hbox{$>$}}}\lower.9ex\hbox{\kern-.190em $\sim$}}}

\newcommand{\CO}{\mathcal{O}}

\newcommand{\VectorTypefaceArrow}{
\let\oldvec\vec
\renewcommand{\vec}[1]{\oldvec{##1}} % use arrow
\newcommand{\uvec}[1]{\hat{##1}} % use a hat
}

\VectorTypefaceArrow

\usepackage{upgreek} % nonslanted greek letters (for unit vectors in spherical coordinates)
\newcommand{\q}{{\widetilde{q}}}

\usepackage{mathtools}

\usepackage{booktabs}
\usepackage{xcolor}
\usepackage{arydshln}

\allowdisplaybreaks

\author[a,b]{Sunghyun Kang,}
\author[a,b]{Injun Jeong,}
\author[a,b]{Stefano Scopel,}
\affiliation[a]{Center for Quantum Spacetime, Sogang University, Seoul 121-742, South Korea}
\affiliation[b]{Department of Physics, Sogang University, Seoul 121-742, South Korea}
\emailAdd{francis735@naver.com}
\emailAdd{natson@naver.com}
\emailAdd{scopel@sogang.ac.kr}

\title{Bracketing the direct detection exclusion plot for a WIMP of spin one half in non--relativistic effective theory }

\abstract{Assuming a standard Maxwellian velocity distribution for the WIMPs in the halo of our Galaxy we use the null results of an exhaustive set of 9 direct detection experiments to calculate the maximal variation of the exclusion plot for each Wilson coefficient of the most general  Galilean--invariant effective Hamiltonian for a WIMP of spin one half due to interferences. We consider 56 Wilson coefficients $c_i^{p,n}$ and $\alpha_i^{n,p}$ for WIMP--proton and WIMP--neutron contact interactions ${\cal O}_i^{p,n}$ and the corresponding long range interaction ${\cal O}_i^{p,n}/q^2$, parameterized by a massless propagator $1/q^2$. For each coupling we provide a different exclusion plot when the following set of operators is allowed to interfere: proton--neutron, i.e. $c_i^{p}$--$c_i^{n}$ or $\alpha_i^{p}$--$\alpha_i^{n}$; contact-contact or long range--long range, i.e. $c_i^{p,n}$--$c_j^{p,n}$ or $\alpha_i^{p,n}$--$\alpha_j^{p,n}$; contact-- long range, i.e. $c_i^{p,n}$--$\alpha_j^{p,n}$. For each of the 56 Wilson coefficients $c_i^{p,n}$ and $\alpha_j^{p,n}$ and for the largest number of interfering operators the exclusion plot variation can reach 3 orders of magnitude and reduces to a factor as small as a few for the Wilson coefficients of the effective interactions where the WIMP couples to the nuclear spin, thanks to the combination of experiments using proton--odd and neutron--odd targets. Some of the conservative bounds require an extremely high level of cancellation, putting into question the reliability of the result. We analyze this issue in a systematic way, showing that it affects some of the couplings driven by the operators  ${\cal O}_{1}$, ${\cal O}_{3}$, ${\cal O}_{11}$, ${\cal O}_{12}$ and ${\cal O}_{15}$, especially when interferences among contact and long range interactions are considered.}

\begin{document}
\hspace*{107.5mm}{CQUeST-2022-0696}\\
\maketitle
\section{Introduction}
\label{sec:introduction}

Weakly Interacting Massive Particles (WIMPs) with a mass in the GeV--TeV range and weak-type interactions with ordinary matter represent the most popular and natural Dark Matter (DM) candidates that are expected to provide the still unaccounted for 27\% of the total mass density of the Universe and more than 90\% of the halo of our Galaxy. Their small but non vanishing interactions can drive WIMP scatterings off nuclear targets, and the measurement of the ensuing nuclear recoils in low-background detectors (direct detection, DD) represents the most straightforward way to detect them (see for instance~\cite{DD_Schumann2019, Snowmass_Leane2022}). 

In absence of a detection, it is customary to represent the null results of DD searches with exclusion plots where the upper bound on the WIMP--nucleon cross section is provided as a function of the WIMP mass $m_\chi$. This procedure is straightforward when the WIMP--proton and WIMP--neutron interactions are fixed to a specific case, and has been applied for a long time in the case of an isospin--conserving  spin--independent (SI) interaction or for a WIMP--proton or WIMP--neutron spin--dependent (SD) coupling. 
On the other hand it is less trivial when no assumption is made on the specific type of interaction that the WIMPs have with neutrons and protons. In such case for a WIMP of spin one half WIMP--nucleus scattering is driven by the most general Galilean--invariant WIMP--nucleon effective Hamiltonian~\cite{nreft_haxton1, nreft_haxton2} up to linear terms on the WIMP velocity:

\begin{equation}
    {\cal H}=\sum_{\tau=0,1}\sum_{j=1}^{15} c_i^\tau {\cal O}_i t^\tau,
    \label{eq:H}
\end{equation}

\noindent in terms of the 15 effective operators listed in Table~\ref{tab:operators}.
In such table $1_{\chi N}$ is the identity operator,
$\vec{q}$ is the transferred momentum, $\vec{S}_{\chi}$ and
$\vec{S}_{N}$ are the WIMP and nucleon spins, respectively, while
$\vec{v}^\perp = \vec{v} + \frac{\vec{q}}{2\mu_{\chi {\cal N}}}$ (with
$\mu_{\chi {\cal N}}$ the WIMP--nucleon reduced mass) is the relative
transverse velocity operator satisfying $\vec{v}^{\perp}\cdot
\vec{q}=0$. In the classification of~\cite{nreft_haxton1, nreft_haxton2}  the operator $\CO_2 = (v^\perp)^2$ is also introduced, but since it is quadratic in the WIMP velocity is not included in the list of Table~\ref{tab:operators}. Moreover, in Eq.~(\ref{eq:H}) $t^0$ = $\mathbb{1}$ , $t^1$ = $\tau^3$ denote the 2 $\times$ 2 identity and third Pauli matrix in isospin space, respectively, and the isoscalar and isovector coupling constants $c^0_j$ and $c^1_j$ are related to those to protons and neutrons $c^p_j$ and $c^n_j$ by $c^0_j$ = $c^p_j$ + $c^n_j$ and $c^1_j$ = $c^p_j$ - $c^n_j$.

\begin{table}[]
\begin{center}
\begin{tabular}{|l|l|l|l|l|l|l|l|}
\hline\hline
\multicolumn{4}{l|}{\multirow{7}{*}{}} & \multicolumn{4}{l}{\multirow{7}{*}{}} \\
\multicolumn{4}{l|}{$ \CO_1 = 1_\chi 1_N$} & \multicolumn{4}{l}{$\CO_9 = i \vec{S}_\chi \cdot (\vec{S}_N \times {\vec{q} \over m_N})$} \\
% \multicolumn{4}{l|}{$\CO_2 = (v^\perp)^2$} & \multicolumn{4}{l}{$\CO_9 = i \vec{S}_\chi \cdot (\vec{S}_N \times {\vec{q} \over m_N})$}\\
\multicolumn{4}{l|}{$\CO_3 = i \vec{S}_N \cdot ({\vec{q} \over m_N} \times \vec{v}^\perp)$} & \multicolumn{4}{l}{$\CO_{10} = i \vec{S}_N \cdot {\vec{q} \over m_N}$} \\
\multicolumn{4}{l|}{$\CO_4 = \vec{S}_\chi \cdot \vec{S}_N$} & \multicolumn{4}{l}{$\CO_{11} = i \vec{S}_\chi \cdot {\vec{q} \over m_N}$} \\
\multicolumn{4}{l|}{$\CO_5 = i \vec{S}_\chi \cdot ({\vec{q} \over m_N} \times \vec{v}^\perp)$} & \multicolumn{4}{l}{$\CO_{12} = \vec{S}_\chi \cdot (\vec{S}_N \times \vec{v}^\perp)$} \\
\multicolumn{4}{l|}{$\CO_6=
  (\vec{S}_\chi \cdot {\vec{q} \over m_N}) (\vec{S}_N \cdot {\vec{q} \over m_N})$} & \multicolumn{4}{l}{$\CO_{13} =i (\vec{S}_\chi \cdot \vec{v}^\perp  ) (  \vec{S}_N \cdot {\vec{q} \over m_N})$} \\
\multicolumn{4}{l|}{$\CO_7 = \vec{S}_N \cdot \vec{v}^\perp$} & \multicolumn{4}{l}{$\CO_{14} = i ( \vec{S}_\chi \cdot {\vec{q} \over m_N})(  \vec{S}_N \cdot \vec{v}^\perp )$} \\
\multicolumn{4}{l|}{$\CO_8 = \vec{S}_\chi \cdot \vec{v}^\perp$} & \multicolumn{4}{l}{$\CO_{15} = - ( \vec{S}_\chi \cdot {\vec{q} \over m_N}) ((\vec{S}_N \times \vec{v}^\perp) \cdot {\vec{q} \over m_N})$} \\ \hline
\end{tabular}
\caption{Non-relativistic Galilean invariant operators for of a WIMP of spin $1/2$ and up to linear terms in the WIMP velocity.}
\label{tab:operators}
\end{center}
\end{table}

When the WIMP--nucleus interaction is driven by the effective Hamiltonian (\ref{eq:H})
a wide parameter space opens up, consisting in 28 independent Wilson coefficients $c_i^{\tau}$ with dimension ${\rm GeV}^{-2}$, that, if assumed as constant, represent the most general contact (short--range) interaction between a WIMP and a nucleon allowed by Galilean invariance.

A first systematic attempt to use the null results from experimental direct detection searches to calculate the exclusion plots on each of the Wilson coefficients $c_i^{\tau}$ of Eq.~(\ref{eq:H}) was first made in~\cite{Catena_Gondolo_2014} and \cite{Catena_Gondolo_2015}, where a global multidimensional statistical analysis was performed to obtain the marginalized posterior probability
density functions (in a Bayesian approach) and the profile likelihoods (in a frequentist approach), as well as associated credible regions and confidence levels. In particular these early analyses showed that the large dimensionality of the parameter space imply several problems. For instance, a Bayesian approach is numerically faster, but the posterior distributions on the single couplings obtained integrating out marginal parameters are affected by large volume effects~\cite{Catena_Gondolo_2014}. On the other hand the profile likelihood in a frequentist approach, where the likelihood function is maximized with respect to the marginal parameters, is not affected by volume effects but is slow and suffers from numerical instabilities~\cite{Catena_Gondolo_2015}. More importantly, in a purely numerical approach it is impossible to assess the quality of the convergence. Indeed, in Ref.~\cite{Catena_Gondolo_2015} it was found that destructive interference effects weaken standard direct detection exclusion limits by up to one order of magnitude in the coupling constants while a semi--analytic approach shows that the effect can be larger~\cite{complementarity_munich_sogang} (see also Section~\ref{sec:discussion}). Crucially, semi--analytic approaches allow to check in a straightforward way if the optimization procedure has converged, so that numerically stable scenarios can be clearly set apart from those that are unstable, allowing for a detailed assessment of the robustness of the method. For this reason semi--analytic approaches represent a useful alternative to purely numerical multidimensional statistical analyses.

 Semi--analytic methods are based on linear algebra, and exploit the fact that within the effective theory of Eq.~(\ref{eq:H}) the expected rate $R$ of a direct detection experiment is a quadratic form in terms of the Wilson coefficients:

\begin{equation}
R=\bm{c}^t\cdot {\cal R}\cdot \bm{c},
\label{eq:ellipsoid}
\end{equation}

\noindent with $\bm{c}$ a 28--dimensional vector containing all the Wilson coefficients and $\cal{R}$ a 28 $\times$ 28 dimensional matrix. As a consequence, the upper bound $N^{max}$ on the count rate ascribable to a WIMP signal that can be obtained by a null search singles up an allowed parameter space that lies inside the multi--dimensional ellipsoid $R<N^{max}$. 

Based on such approach a consistent proposal on how to perform a generalization of the concept of exclusion plot was recently put forward in Refs.~\cite{conservative_icecube_2021, complementarity_munich_sogang}.  The strategy adopted in such papers is to bracket the maximal variation of the exclusion plot on each Wilson coefficient due to the interference with all the other couplings. In such approach the exclusion plot of a given $c_i^\tau$ is no longer a line, but generalizes to a band delimited by the most constraining bound, which is obtained  by assuming that $c_i^\tau$ is the only non--vanishing coupling, and the less constraining one, which corresponds to the maximal cancellation among the contribution of $c_i^\tau$ and that of all the other couplings of the effective theory.
\begin{figure}[h]
\centering
\begin{tabular}{cc}
\includegraphics[width=8cm]{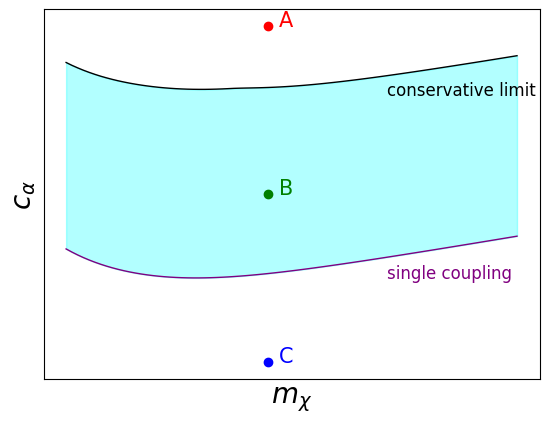} 
\end{tabular}
\caption{Exemplification of the meaning of the exclusion bands discussed in Section~\ref{sec:exclusion_bands}. Point A is certainly ruled out, no matter which
ultraviolet completion it is generated from; point C is certainly allowed, again no matter which ultraviolet completion it is generated from; Point B can be excluded or not, depending on the specific model at high energy it is generated from.
\label{fig:example_bands}}
\end{figure}
In particular, if the numerical value of a coupling exceeds the corresponding most conservative constraint it is certainly ruled out, no matter which ultraviolet completion it is generated from (point A in Fig.~\ref{fig:example_bands}); if a coupling is below the most constraining bound it is possible to
conclude that the value is allowed, again no matter which ultraviolet completion it is generated from (point C in Fig.~\ref{fig:example_bands}). On the other hand, if its value is within the exclusion band (point B in Fig.~\ref{fig:example_bands}) it could be excluded or not, depending on the specific model at high energy it is generated from. This can be useful to constrain a wide class of theoretical models without the need to re--analyze the experimental data, since the exclusion plots published by experimental collaborations are only valid in specific scenarios.

The method introduced in Refs.~\cite{conservative_icecube_2021, complementarity_munich_sogang} can handle experiments with several energy bins (a different ellipse for each bin) but cannot be applied to unbinned analyses. Another issue of the method is that at low WIMP masses it shows numerical instabilities due to the fact that all signals are suppressed by the tail of the velocity distribution and become very sensitive to the parameters. As a consequence, it is not suitable to obtain conservative bounds for light WIMPs.

The procedure of Ref.~\cite{conservative_icecube_2021} determines a different direction of maximal cancellation for each bound (i.e. for each matrix ${\cal R}$). 
Such approach can be seen as a multi--dimensional generalization of what was done in~\cite{feng_isospin_dm_2011} for a spin--independent interaction , where the conservative constraint 
on the WIMP--nucleus cross section $\sigma_{\chi N}\propto [c^p Z+c^n (A-Z)]^2$ (with $Z$ the atomic number and $A$ the mass number) was obtained by tuning the ratio between the WIMP--proton and the WIMP--neutron couplings $c_n/c_p=Z/(Z-A)$ for the target of a given experiment. In particular, the simple prescription suggested in~\cite{feng_isospin_dm_2011} consisted in determining the direction of maximal cancellation in the two--dimensional plane of $c_n$ and $c_p$. In~\cite{conservative_icecube_2021} the same concept was generalized to the case of a signal of the form (\ref{eq:ellipsoid}) by providing a semi--analytical procedure to determine an analogous direction of maximal cancellation for a given target in the multi--dimensional vector space of the $c_i^\tau$ Wilson coefficients. 

In~\cite{complementarity_munich_sogang} it was found that such procedure can lead to instabilities due to flat directions in ${\cal R}$ when the matrices are close to singular. In Section~\ref{app:factorization} we will give some semi--analytical arguments to explain why this indeed may happen. Moreover, also in~ \cite{complementarity_munich_sogang} such procedure was generalized to combine the bounds from different DD experiments, i.e. to find the direction of maximal cancellation by combining several matrices ${\cal R}_i$.  
Crucially, the outcome of such procedure consists in the determination of intersections of ellipsoids and hyperplanes (see Section~\ref{sec:conservative}) so that, at variance with multidimensional likelihood analyses, it is straightforward to assess the convergence of the optimization process by verifying that the solution corresponds to the desired intersection (see for instance Figs.~\ref{fig:c1_ellipsoids} and \ref{fig:c1_full_ellipsoids} and corresponding captions).

Ref.~\cite{complementarity_munich_sogang} was devoted to introducing the method, and provided the quantitative calculation of the conservative exclusion plots by including two direct detection experiments (XENON1T~\cite{xenon_2018} and PICO--60($C_3F_8$)~\cite{pico60_2019}) for a contact interaction. Moreover, in~\cite{complementarity_munich_sogang} the bounds from DD were combined with those from dark matter capture in the solar interior, resulting from the non observation
of a neutrino excess in the IceCube data collected in the direction of the Sun~\cite{Gould_1987_capture_in_the_sun}. 

In the present paper we wish to make a first quantitative and systematic assessment of the impact and effectiveness of the semi--analytical approach summarized above for DD. In particular, with the goal of obtaining conservative bounds, we do not include WIMP capture in the Sun, which requires additional assumptions besides the effective Hamiltonian in~(\ref{eq:H})~\footnote{In~\cite{complementarity_munich_sogang} equilibrium between capture and annihilation was assumed, as well as annihilation into $W^+W^-$ for $m_\chi>$  100 GeV and $\tau^+\tau^-$ for $m_\chi<$ 100 GeV.} 

In particular we extend the analysis of Ref.~\cite{complementarity_munich_sogang} in  two directions. First, we calculate the conservative bounds on each of the Wilson coefficients by extending the list of direct detection experiments (besides XENON1T and PICO--60($C_3F_8$) we include LZ~\cite{LZ_2022}, PandaX--4T~\cite{pandax4T_2021}, PICO--60 ($CF_3I$)~\cite{pico60_2015}, SuperCDMS~\cite{super_cdms_2017},  CDMSlite~\cite{cdmslite_2017}, COSINE--100~\cite{cosine_bck} and DAMIC~\cite{DAMIC_2016}). Moreover, we generalize the Hamiltonian of Eq.~(\ref{eq:H}) with:

\begin{equation}
    {\cal H}=\sum_{\tau=0,1}\sum_{j=1}^{15} \left(c_i^\tau+\frac{\alpha_i^\tau}{q^2} \right) {\cal O}_i t^\tau=\sum_{\tau=0,1}\sum_{j=1}^{15} c_i^\tau {\cal O}_i t^\tau+\sum_{\tau=0,1}\sum_{j=1}^{15} \alpha_i^\tau \frac{{\cal O}_i}{q^2} t^\tau
    \label{eq:H_short_long}.
\end{equation}

\noindent to include for each operator ${\cal O}_i$ the effect of a long range interaction parameterized by the momentum dependence of a massless propagator $1/q^2$.
The very general parameterization of the
DM scattering process of Eq.~(\ref{eq:H_short_long}) captures almost all conceivable particle physics scenarios for the interaction of DM with nucleons~\footnote{The only exception is provided by interactions that induce a meson pole with scaling $1/(m_k - q^2)$ with $k$ = $\pi$, $\eta$~\cite{Bishara_2017}.}  and depends on 
56 independent Wilson coefficients $c_i^{\tau}$, $\alpha_i^{\tau}$ (in the isospin base) or $c_i^{p,n}$, $\alpha_i^{p,n}$ (in the proton--neutron base).

The main quantitative results of our paper are contained in Figs. \ref{fig:SI_shade}, \ref{fig:SD_shade}, \ref{fig:no_op_interf_shade}, \ref{fig:SI_long_shade}, \ref{fig:SD_long_shade}, \ref{fig:no_op_interf_long_shade}.
Such figures contain a total of 152 exclusion bands, that represent the maximal variation of the exclusion plot of each coupling when the latter is allowed to interfere with different sets of operators. They provide the first quantitative and systematic discussion of a model--independent generalization of the direct detection exclusion plot for a WIMP of spin one half. As already pointed out, in some cases a high sensitivity of the result, related to the level of cancellation required by the conservative bounds, is observed on the input values of the matrices ${\cal R}$. We analyze this issue in a systematic way in Section~\ref{sec:cancellations}, where we show that, indeed, for some of the couplings the reliability of the conservative upper bounds discussed in Section~\ref{sec:exclusion_bands}  should be considered with care, given the level of numerical accuracy of the calculation.  

All the quantitative results of the paper have been obtained using the \verb|WimPyDD|~\cite{wimpydd_2022} code. In particular a new version of \verb|WimPyDD| was released containing the routine \verb|wimp_dd_| \verb|matrix| that allows to calculates in a straightforward way the matrix ${\cal R}$ for a generic set of operators and a given experimental setup and including the response of the detector (see Appendix~\ref{app:wimpydd}).   

The plan of the paper is the following: in Section~\ref{sec:expected_rates}
we outline the main ingredients needed to calculate the expected rate in a DD experiment within an effective theory described by the Hamiltonian of Eq.~(\ref{eq:H_short_long}); in Section~\ref{sec:conservative} we outline the procedure to obtain conservative bounds including interferences among different operators, both in the case of a single experiment and for the combination of different ones. In Section~\ref{sec:exclusion_bands} we discuss our results in terms of exclusion bands for each Wilson coefficient, and in Section~\ref{sec:cancellations} we focus on the issue of cancellations. Section~\ref{sec:conclusions} is devoted to our conclusions. In Appendix~\ref{app:factorization} we explicitly show how the scattering amplitude at fixed momentum transfer can be written as the sum  of squares of polynomials linear in the Wilson coefficients, providing some insight on why the matrices ${\cal R}$ can be close to singular for some experiments; in Appendix~\ref{app:experiments} we provide details on how the experimental bounds were implemented; finally, in Appendix~\ref{app:wimpydd} we introduce the routine \verb|wimp_dd_matrix| included in the new release of \verb|WimPyDD|,
that calculates the matrices ${\cal R}$ for a given effective Hamiltonian and experimental set--up.

\section{Expected rate for WIMP-nucleus scattering}
\label{sec:expected_rates}

In this section we summarize the expressions for the calculation of the WIMP--nucleon elastic scattering in non--relativistic effective theories. 
More details can be found for instance in~\cite{nreft_sensitivity_sogang,eft_relativistic_sogang_2020,wimpydd_2022}.

The expected rate in a given visible energy bin $E_1^{\prime}\le
E^{\prime}\le E_2^{\prime}$ of a direct detection experiment is given
by:

\begin{eqnarray}
R_{[E_1^{\prime},E_2^{\prime}]}&=&M\mbox{T}\int_{E_1^{\prime}}^{E_2^{\prime}}\frac{dR}{d
  E^{\prime}}\, dE^{\prime}, \label{eq:start}\\
 \frac{dR}{d E^{\prime}}&=&\sum_T\left(\frac{dR}{d E^{\prime}}\right )_T=\sum_T \int_0^{\infty} \frac{dR_{\chi T}}{dE_{ee}}{\cal
   G}_T(E^{\prime},E_{ee})\epsilon(E^{\prime})\label{eq:start2}\,d E_{ee}, \label{eq:diff_rate_eprime}\\
E_{ee}&=&q(E_R) E_R \label{eq:start3},
\end{eqnarray}

\noindent with $\epsilon(E^{\prime})\le 1$ the experimental
efficiency/acceptance. In the equations above $E_R$ is the recoil
energy deposited in the scattering process (indicated in keVnr), while
$E_{ee}$ (indicated in keVee) is the fraction of $E_R$ that goes into
the experimentally detected process (ionization, scintillation, heat)
and $q(E_R)$ is the quenching factor, ${\cal
  G_T}(E^{\prime},E_{ee}=q(E_R)E_R)$ is the probability that the
visible energy $E^{\prime}$ is detected when a WIMP has scattered off
an isotope $T$ in the detector target with recoil energy $E_R$, $M$ is
the fiducial mass of the detector and T the live--time of the data
taking. For a given recoil energy imparted to the target the
differential rate for the WIMP--nucleus scattering process is given
by:

\be
\frac{d R_{\chi T}}{d E_R}(t)=\sum_T N_T\frac{\rho_\chi}{m_\chi}\int_{v_{min}}d^3 v_T f(\vec{v}_T,t) v_T \frac{d\sigma_T}{d E_R},
\label{eq:dr_de}
\ee

\noindent where $N_T$ is the number of the nuclear targets of
species $T$ in the detector (the sum over $T$ applies in the case of
more than one nuclear isotope), $\rho_{\chi}$ is the local WIMP mass density in the neighborhood of the Sun, $f(\vec{v}_T)$ is the WIMP velocity distribution, for which
we assume a standard isotropic Maxwellian at rest in the Galactic rest
frame with velocity dispersion $v_{rms}$=270 km/s truncated at the galactic escape velocity $u_{esc}$=550 km/s, and boosted to the Lab frame by the velocity of the Solar system $v_0$=220 km/s. Moreover:
\begin{equation}
v_{min}^2=\frac{q^2}{4 \mu_{T}^2}=\frac{m_T E_R}{2 \mu_{T}^2},
\label{eq:vmin}
\end{equation}
\noindent represents the minimal incoming WIMP speed required to
impart the nuclear recoil energy $E_R$ (with $(v^{\perp})^2$ = $v^2-v_{min}^2$) and:
\be
\frac{d\sigma_T}{d E_R}=\frac{2 m_T}{4\pi v_T^2}\left [ \frac{1}{2 j_{\chi}+1} \frac{1}{2 j_{T}+1}|\mathcal{M}_T|^2 \right ].
\label{eq:dsigma_de}
\ee

\noindent Assuming that the nuclear interaction is the sum of the
interactions of the WIMPs with the individual nucleons in the nucleus (one--nucleon approximation) one has:

\begin{equation}
  \frac{1}{2 j_{\chi}+1} \frac{1}{2 j_{T}+1}|\mathcal{M}_T|^2=
  \frac{4\pi}{2 j_{T}+1} \sum_{\tau=0,1}\sum_{\tau^{\prime}=0,1}\sum_{k} R_k^{\tau\tau^{\prime}}\left [c^{\tau}_i, \alpha^\tau_i,(v^{\perp}_T)^2,\frac{q^2}{m_N^2}\right ] W_{T k}^{\tau\tau^{\prime}}(y).
\label{eq:squared_amplitude}
\end{equation}

\noindent In the above expression $j_{\chi}$ and $j_{T}$ are the WIMP
and the target nucleus spins, respectively, $q=|\vec{q}|$ while the
$R_k^{\tau\tau^{\prime}}$'s are WIMP response functions (that we
report for completeness in Eq.(\ref{eq:wimp_response_functions}))
which depend on the couplings $c^{\tau}_i$, $\alpha^\tau_i$ as well as the transferred momentum $\vec{q}$ and $(v^{\perp}_T)^2$, and that can be decomposed in a velocity--independent and a velocity--dependent part:

\begin{equation}
R_k^{\tau\tau^{\prime}}=R_{0k}^{\tau\tau^{\prime}}+R_{1k}^{\tau\tau^{\prime}} (v^{\perp}_T)^2=R_{0k}^{\tau\tau^{\prime}}+R_{1k}^{\tau\tau^{\prime}}\left (v_T^2-v_{min}^2\right ).
\label{eq:r_decomposition}
\end{equation}

Moreover, in equation
(\ref{eq:squared_amplitude}) the $W^{\tau\tau^{\prime}}_{T k}(y)$'s
are nuclear response functions and the index $k$ represents different
effective nuclear operators, which, crucially, under the assumption
that the nuclear ground state is an approximate eigenstate of $P$ and
$CP$, can be at most eight: following the notation in
\cite{nreft_haxton1,nreft_haxton2}, $k$=$M$, $\Phi^{\prime\prime}$,
$\Phi^{\prime\prime}M$, $\tilde{\Phi}^{\prime}$,
$\Sigma^{\prime\prime}$, $\Sigma^{\prime}$,
$\Delta$,$\Delta\Sigma^{\prime}$. 

In the analysis of Section~\ref{sec:discussion} we will use null results from DD experiments to put upper bounds on the quantity:

\begin{equation}
  R_{[E_1^{\prime},E_2^{\prime}]}=MT\sum_T\int_{E_1^{\prime}}^{E_2^{\prime}}d E^{\prime}\;\left(\frac{dR}{d E^{\prime}} \right)_T
  \label{eq:start1},
\end{equation}

\noindent which represents the expected number of events in a WIMP DD experiment in the interval of visible energy $E_1^{\prime}\le E^{\prime}\le
E_2^{\prime}$. As evident from Eq.~(\ref{eq:wimp_response_functions}) the $R_k^{\tau\tau^{\prime}}$ response functions, and so also $R_{[E_1^{\prime},E_2^{\prime}]}$ are quadratic forms in the couplings $c^{\tau}_i$, $\alpha^\tau_i$, so that, as anticipated in Eq.~(\ref{eq:ellipsoid}), an upper bound on $R_{[E_1^{\prime},E_2^{\prime}]}$ determines an allowed region inside an ellipsoid.   

The $W^{\tau\tau^{\prime}}_{T k}(y)$'s are function of $y\equiv (qb/2)^2$, where $b$ is the size of the nucleus. For the target nuclei $T$ used in most direct detection experiments the functions $W^{\tau\tau^{\prime}}_{T k}(y)$, calculated using nuclear shell models, have been provided in
Refs.~\cite{nreft_haxton2,Catena_nuclear_form_factors}.
Such calculations are at the Born level under the assumption that the dark matter particle couples to the nucleus through local one--body interactions with the nucleons.  %Notice that two--body effects~\cite{Cirigliano_two_bodies_2012,Cirigliano_two_bodies_2014,Klos_two_bodies_2013,Vietze_two_bodies_2014}, which are only available for a few isotopes, can be important when the one--body contribution is suppressed. However minimizing the expected rate through maximizing the level of cancellation among the contributions of different couplings is precisely the procedure that we will use to obtain the conservative bounds of Section~\ref{sec:discussion}. This aspect will warrant a closer inspection in the next Section.

\section{Conservative bounds}
\label{sec:conservative}

The procedure to obtain a conservative bound on the coupling $c_\alpha$ is exemplified in Fig.~\ref{fig:ellipses_bracketing} for the case of a two--dimensional parameter space, where $c_\alpha$ interferes with the coupling $c_\beta$.
In such figures the regions inside the two ellipses represent the parameter space allowed by the constraints of two different experiments Exp 1 and Exp 2, represented by the two red-- and green-shaded areas, respectively:

\begin{eqnarray}
R_{Exp 1}&=&\bm{c}^t\cdot {\cal R}_{Exp\, 1}\cdot \bm{c}=\sum_{i,j=\alpha,\beta} c_i ({\cal R}_{Exp\,1})_{ij}c_j<1, \nonumber\\
R_{Exp 2}&=&\bm{c}^t\cdot {\cal R}_{Exp\, 2}\cdot \bm{c}=\sum_{i,j=\alpha,\beta} c_i ({\cal R}_{Exp\,2})_{ij} c_j<1. 
\end{eqnarray}

\noindent In the equations above and in the following we normalize the matrices of expected rates to the corresponding upper bound, i.e. ${\cal R}_{Exp\, i}\rightarrow{\cal R}_{Exp\,i}/N^{max}_{Exp\, i}$.

\subsection{The case of a single DD constraint}
\label{sec:single_bound}
The case of a single constraint/ellipsoid was discussed in Ref.~\cite{conservative_icecube_2021}. 
When $c_\alpha$ is the only non--vanishing coupling ($c_\beta$ =0) the upper bound from Exp 1 is determined by the intersection of the red ellipse with the $c_\alpha$ axis, i.e. $max(c_\alpha)_{1,c_\beta =0}$ = $\sqrt{1/({\cal R}_{Exp\,1})_{\alpha\alpha}}$. However, when $c_\alpha$ is allowed to interfere with $c_\beta$ its maximal allowed value can be larger, since it is given by the projection on the $c_\alpha$ axis of the point of the red ellipse with maximal distance from the $c_\beta$ axis (dashed blue line). The explicit value of such upper bound is given by~\cite{conservative_icecube_2021}:

\begin{equation}
    max(c_\alpha)_1=\sqrt{({\cal R}_{Exp\,1}^{-1})_{\alpha\alpha}},
    \label{eq:c_max_single_exp}
\end{equation}
\noindent with ${\cal R}_{Exp\,1}^{-1}$ the inverse of the matrix ${\cal R}_{Exp\,1}$.
Clearly, such limit diverges if ${\cal R}_{Exp\,1}$ is singular. Indeed, in Ref.~\cite{complementarity_munich_sogang} it was observed that in some cases the matrices ${\cal R}$ are close to singular, leading to a very large sensitivity of the ensuing bounds on their exact entries. In Appendix~\ref{app:factorization} we discuss the reasons of such numerical aspect in some detail, with the help of analytical expressions.

\subsection{Combining different experiments}
\label{sec:combining}

\begin{figure}[h]
\centering
\begin{tabular}{cc}
\includegraphics[width=12cm]{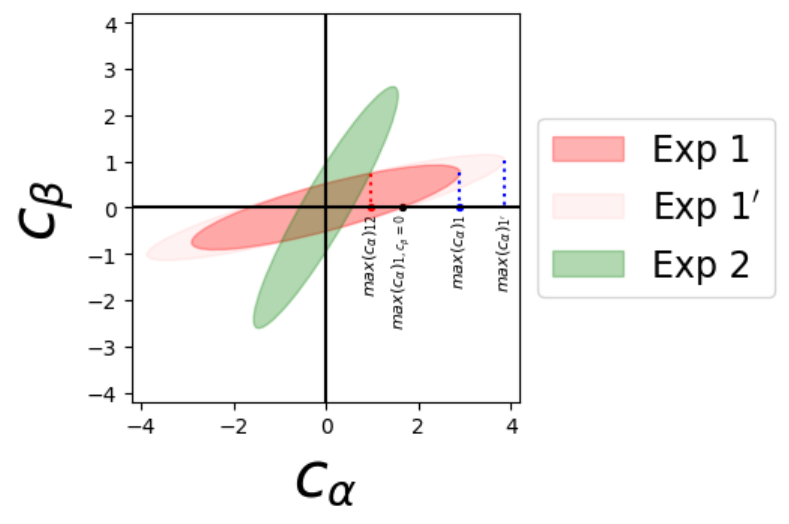} 
\end{tabular}
\caption{Exemplification of the regions of parameter space allowed by the two experiments Exp 1 (red ellipse) and Exp 2 (green ellipse). $\max(c_\alpha)_{1,c_\beta =0}$ (black dot) is the upper bound on $c_\alpha$ from Exp 1 when $c_\alpha$ is the only non--vanishing coupling, and corresponds to the intersection of the ellipse of Exp 1 with the $c_\alpha$ axis; $\max(c_\alpha)_1$ is the bound on $c_\alpha$ from Exp 1 when interference between $c_\alpha$ and $c_\beta$ is allowed, and corresponds to the projection on the $c_\alpha$ axis of the tip of the ellipsoid of Exp 1; finally, $\max(c_\alpha)_{12}$ is the combined limit of Exp 1 and Exp 2 when interference between $c_\alpha$ and $c_\beta$ is allowed, and corresponds to the projection on the $c_\alpha$ axis of the tip of the  combined allowed region given by the overlapping of the ellipses for Exp 1 and Exp 2. Exp 1$^{\prime}$ represents the modification of the ellipse of Exp 1 if the corresponding matrix has a near--vanishing eigenvalue and is sensitive to small perturbations of its entries. In this case the bound from only Exp 1 changes from $\max(c_\alpha)_1$ to $\max(c_\alpha)_{1^{\prime}}$. However $\max(c_\alpha)_{12}$, the combined bound from Exp 1 and Exp 2, is not affected by the sensitivity. \label{fig:ellipses_bracketing}}
\end{figure}

The calculation of a conservative bound on some coupling $c_\alpha$ when the constraints from two different DD experiments Exp 1 and Exp 2 are considered is again exemplified in Fig.~\ref{fig:ellipses_bracketing}. In this case the effect of the combination of the two experiments is to reduce the allowed region to the overlapping of the two ellipsoids. The conservative bound $max(c_\alpha)_{12}$ is then the projection along the $\alpha$ axis of the intersection between the two ellipses (dashed red vertical line). 
%As we will see in Section~\ref{sec:discussion} if the angle between the directions of the semi--major axes of the two ellipses (including the pathological case of almost flat directions discussed in Section~\ref{sec:sensitivity}) is large enough (i.e. if the two experiments are {\it complementary}) the conservative upper bound $max(c_\alpha)$ can be relaxed by a surprisingly low factor compared to the strongest bound obtained assuming a single non--vanishing coupling (see Figs.~\ref{fig:relaxing_factor_20}, \ref{fig:relaxing_factor_100}, \ref{fig:relaxing_factor_1000}).   

%In Ref.~\cite{complementarity_munich_sogang} a procedure to calculate $max(c_\alpha)_{12}$ was introduced based on a constrained maximization procedure. In this procedure $c_\alpha$ is maximized 
%imposing the constraint ${\cal L}-{\cal L}_{min}<n^2$ corresponding to $n$ sigmas for the combined Likelihood function of the experiments ${\cal L}=\sum_i{\cal L}_{Exp\,i}$. The advantage of this approach is that the constrained maximization leading to $max(c_\alpha)_{12}$ requires a single Lagrange multiplier for an arbitrary number of experimental bounds.

\begin{figure}[h]
\centering
\begin{tabular}{cc}
\includegraphics[width=10cm]{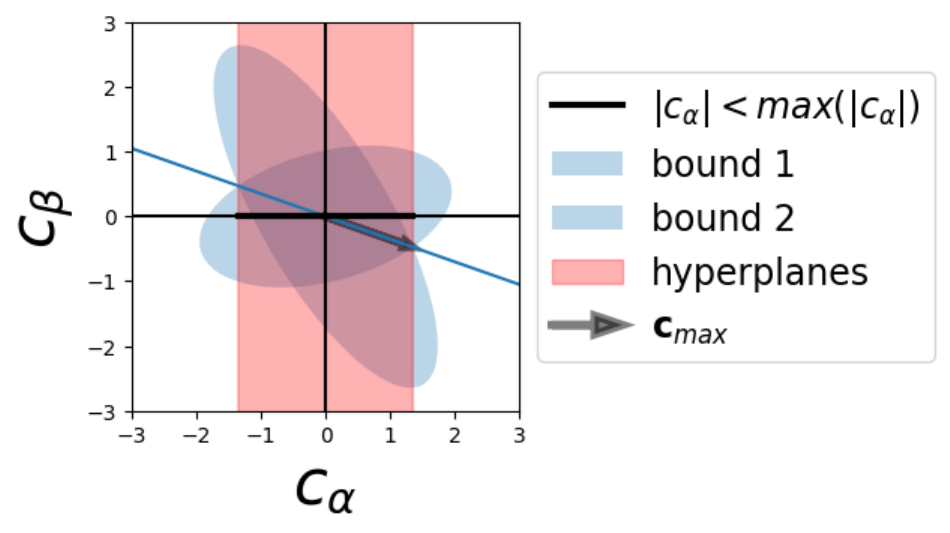} 
\end{tabular}
\caption{The maximal excursion of $c_\alpha$ is found by projecting the allowed region (determined by the ellipsoid intersection) on the $\alpha$ axis. This requires to find the two hyperplanes $|c_\alpha|< max(|c_\alpha|)$ that intersect the tip of the allowed region.
\label{fig:ellipses_bracketing_LMI}}
\end{figure}

The maximal excursion allowed to the coupling $c_\alpha$ is given by the projection of the region of the allowed parameter space on the $\alpha$ axis, as exemplified in Fig.~\ref{fig:ellipses_bracketing_LMI}.
This is a standard elliptic problem that can be solved using available packages such as PICOS~\cite{picos}. In particular, the family of hyperplanes perpendicular to the $\alpha$ axis is given by:

\begin{equation}
  \bm{c}^t\cdot B \cdot\bm{c}\le (\max (c_\alpha))^2,\;\;\;  B=\begin{pmatrix}
0 & 0 & 0 & 0 &\mbox{...}\\
0 & 0 & 0 & 0 & \mbox{...}\\
0 & 0 & 1 & 0 & \mbox{...}\\
0 & 0 & 0 & 0 & \mbox{...}\\
\end{pmatrix}
\label{eq:b_matrix}
\end{equation}

\noindent where the only non--vanishing term on the diagonal of $B$ corresponds to the $c_\alpha$ axis.
Indicating with $A_k$ the matrices of $n$ bounds the value of $max(c_\alpha)$ is the maximal one for which~\cite{s_lemma}:

\begin{eqnarray}
 && \xi_i\ge   0,\,\,\,\sum_{i=1}^{n}  \xi_i\le 1, \label{eq:csi_lmi}\\
  && \sum_{i=1}^{n}  \xi_i A_k-\frac{B}{max(c_\alpha)^2}\,\,\,\mbox{is a positive matrix}.
  \label{eq:lmi}
\end{eqnarray}

\noindent with $\xi_i$ some Lagrange multipliers. In practice the above procedure consists in calculating at fixed values of the $\xi_i$'s the value of $max(c_\alpha)^2$ for which the minimum eigenvalue of the matrix in Eq.~(\ref{eq:lmi}) vanishes with all the other eigenvalues positive, and to maximize $max(c_\alpha)$ as a function of $\xi_i$ with the conditions~(\ref{eq:csi_lmi}). The algorithm of Eqs.~(\ref{eq:csi_lmi},\ref{eq:lmi}) has been already used to study the multi--dimensional parameter space of WIMP--nucleus non--relativistic effective theory in several papers~\cite{Catena_dama_ellipsoids_2016, Catena_cdms_si_ellipsoids_2018, dama_inelastic_2019}. 

In Fig.~\ref{fig:ellipses_bracketing} Exp 1$^{\prime}$ represents the modification of the ellipse Exp 1 under a small perturbation of the corresponding matrix if the latter has a near--vanishing eigenvalue. In this case the corresponding axis of the ellipsoid is sensitive to small perturbations of the matrix input values, and the bound $\max(c_\alpha)_1$ given by Eq.~(\ref{eq:c_max_single_exp}) is unstable. However, $\max(c_\alpha)_{12}$ is not affected. This is at the base of why combining the bounds of different experiments is not only useful to get better constraints, but may also prove to be crucial to get robust results.

\section{Discussion}
\label{sec:discussion}

In this Section we apply the method outlined in \ref{sec:combining} to calculate systematically the conservative upper bound for each of the 56 couplings $c^{p(n)}_1$, $c^{p(n)}_3$,...$c^{p(n)}_{15}$, $\alpha^{p(n)}_1$, $\alpha^{p(n)}_3$,...$\alpha^{p(n)}_{15}$. In our analysis we include the set of 9 experimental bounds listed in Appendix~\ref{app:experiments}: LZ~\cite{LZ_2022}, PandaX--4T~\cite{pandax4T_2021}, XENON1T~\cite{xenon_2018}, PICO--60 ($C_3F_8$)~\cite{pico60_2019}, PICO--60 ($CF_3I$)~\cite{pico60_2015}, SuperCDMS~\cite{super_cdms_2017}, CDMSlite~\cite{cdmslite_2017}, COSINE--100~\cite{cosine_bck} and DAMIC~\cite{DAMIC_2016}. Our results for the variation of the exclusion plot due to interferences (exclusion bands) are discussed in Section~\ref{sec:exclusion_bands}, while the issue of their sensitivity to the input matrices due to large cancellations is addressed in a quantitative and systematic way in Section~\ref{sec:cancellations}. 

\subsection{Exclusion plots}
\label{sec:exclusion_bands}

\begin{figure}[h]
\centering
\begin{tabular}{cc}
\includegraphics[width=10cm]{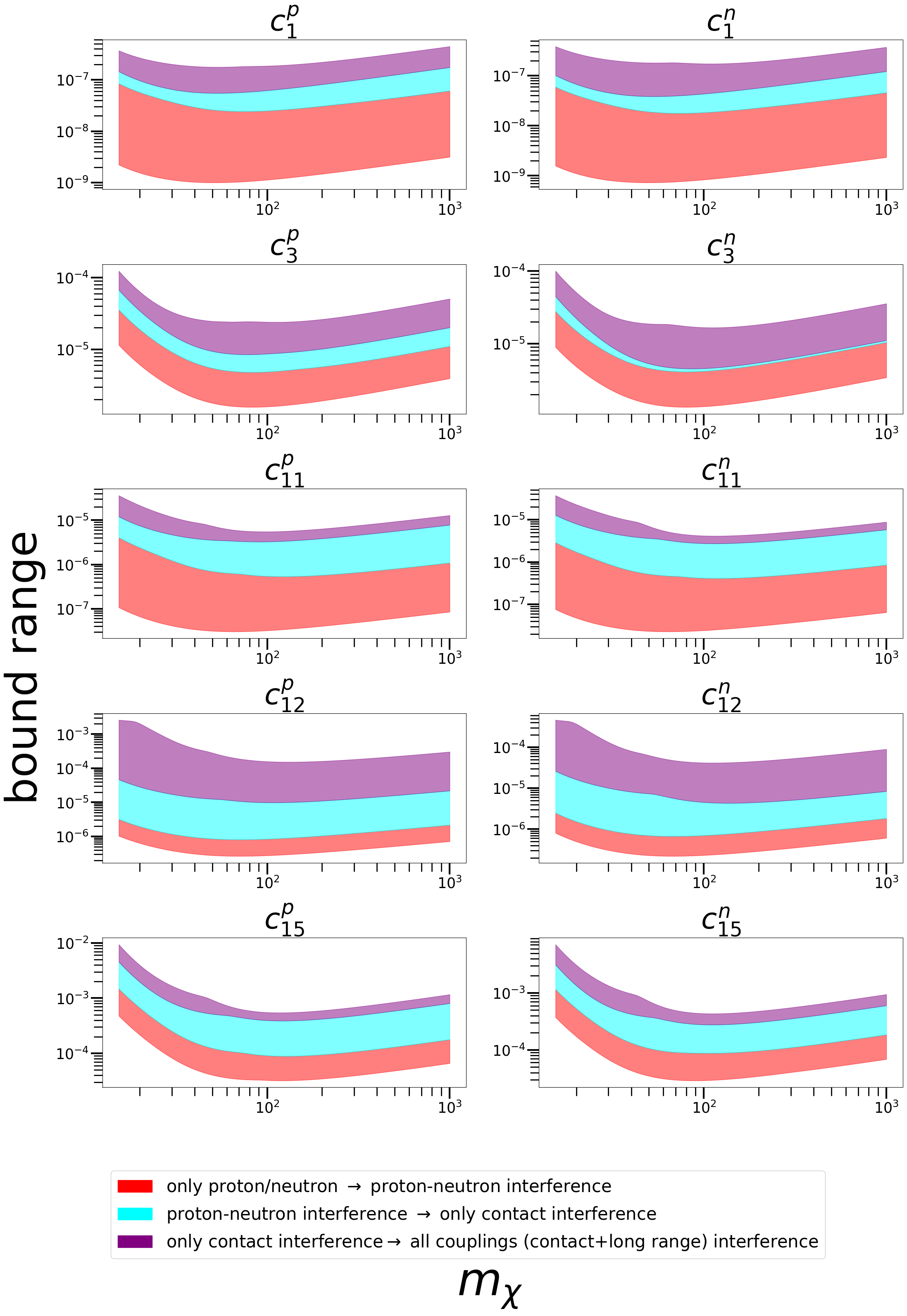} 
\end{tabular}
\caption{Exclusion bands for the "spin--independent" contact WIMP--proton couplings $c^p_{1}$, $c^p_{3}$, $c^p_{11}$, $c^p_{12}$ and $c^p_{15}$ (left column) and the WIMP--neutron couplings $c^n_{1}$, $c^n_{3}$, $c^n_{11}$, $c^n_{12}$ and $c^n_{15}$ (left column) as a function of the WIMP mass $m_\chi$. The lower (red) band is spanned when each operator interferes in its own proton--neutron parameter space;  the middle (cyan) band represents how the variation of the exclusion plot is extended when each operator interferes also with other contact couplings; the upper (purple) band is the additional extension of the exclusion band when interferences between contact and long range interactions is also allowed. \label{fig:SI_shade}}
\end{figure}

\begin{figure}[h]
\centering
\begin{tabular}{cc}
\includegraphics[width=10cm]{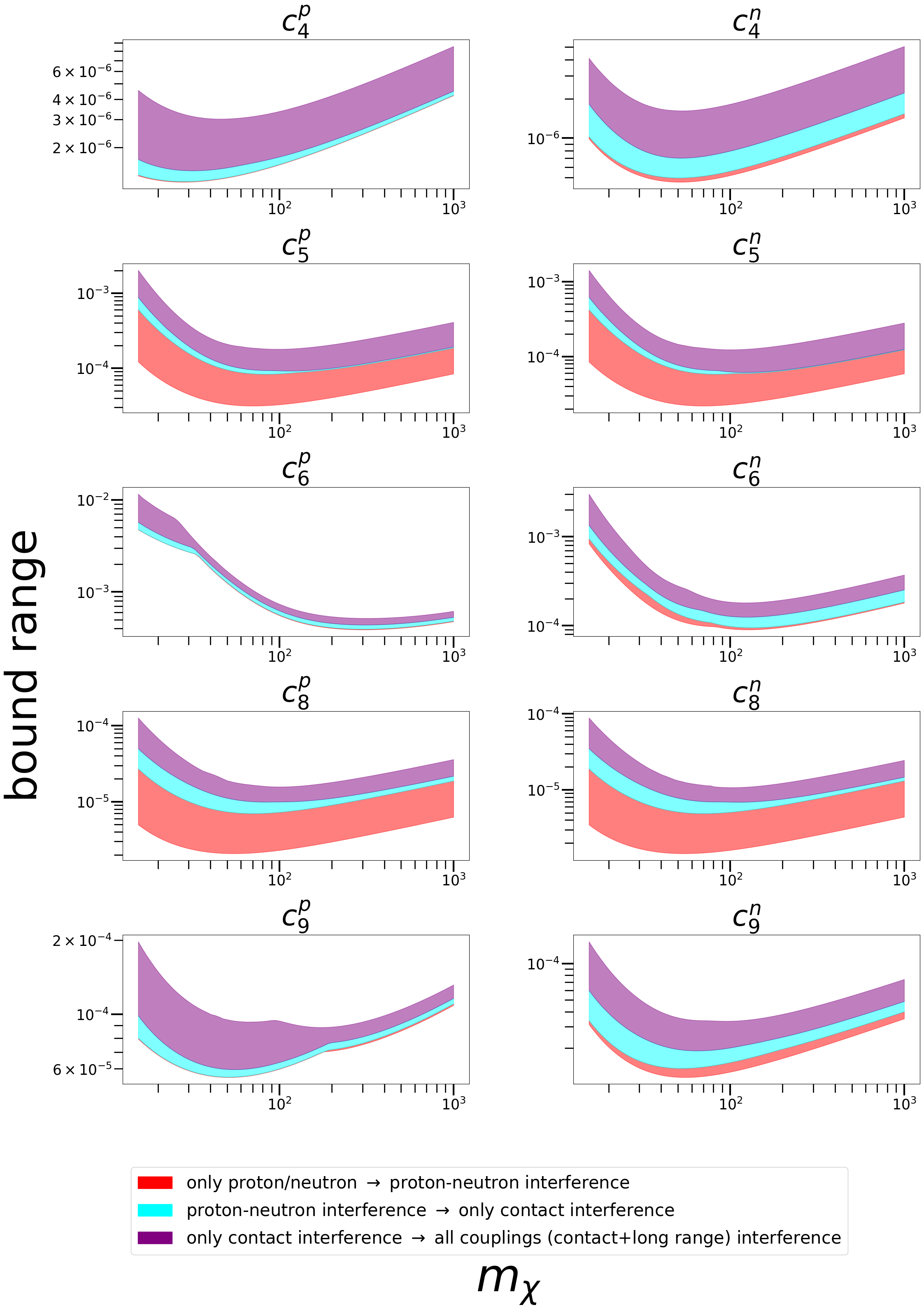} 
\end{tabular}
\caption{The same as in Fig.~\ref{fig:SI_shade} for the "spin--dependent" contact WIMP--proton couplings $c^p_{4}$, $c^p_{5}$, $c^p_{6}$, $c^p_{8}$, $c^p_{9}$ (left column) and the WIMP--neutron couplings $c^n_{4}$, $c^n_{5}$, $c^n_{6}$, $c^n_{8}$, $c^n_{9}$ (right column). \label{fig:SD_shade}}
\end{figure}

\begin{figure}[h]
\centering
\begin{tabular}{cc}
\includegraphics[width=10cm]{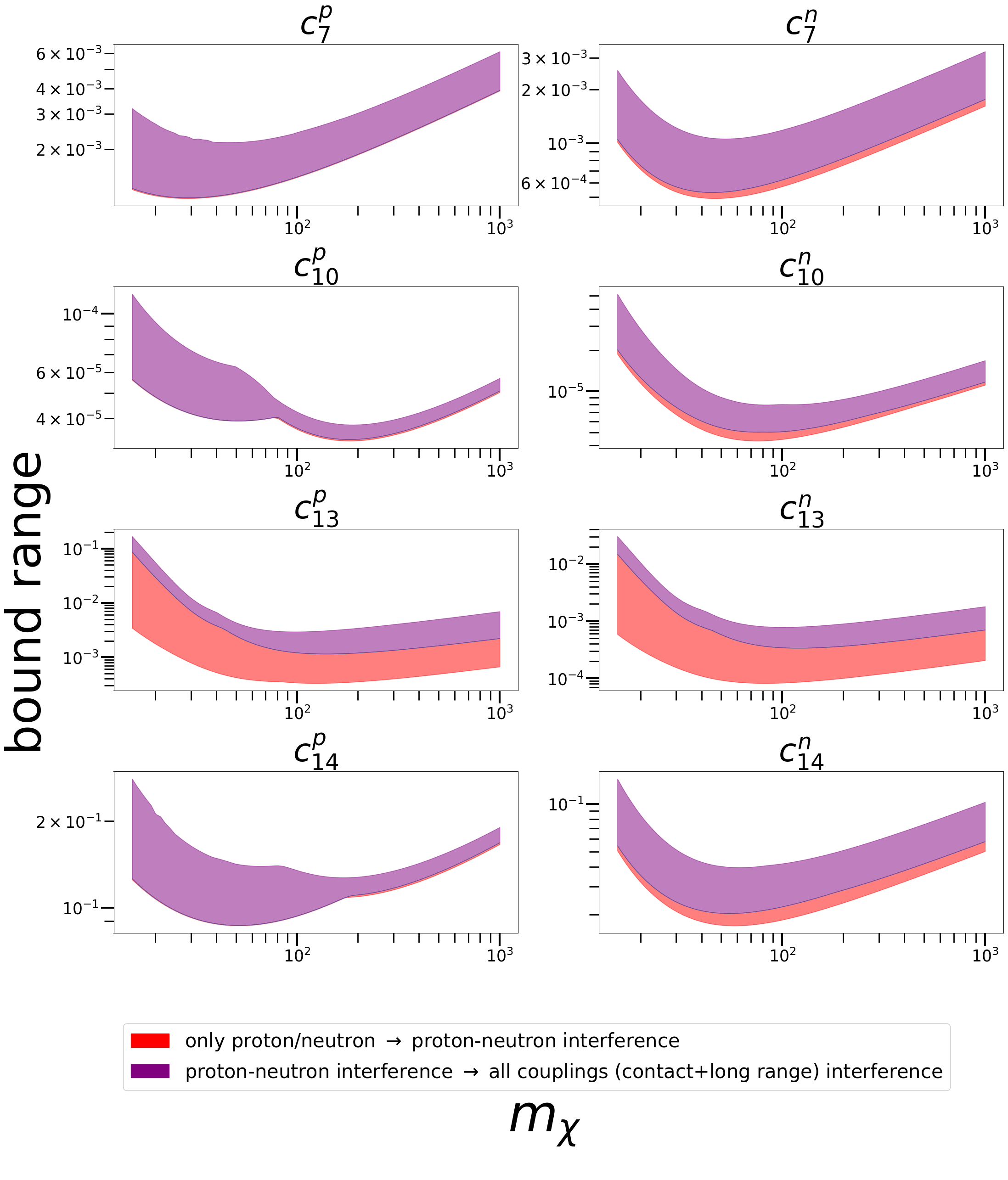} 
\end{tabular}
\caption{The same as in Fig.~\ref{fig:SI_shade} for the non-interfering contact WIMP--proton couplings $c^p_{7}$, $c^p_{10}$, $c^p_{13}$, $c^p_{14}$ (left column) and the WIMP--neutron couplings $c^n_{7}$, $c^n_{10}$, $c^n_{13}$, $c^n_{14}$ (right column). \label{fig:no_op_interf_shade}}
\end{figure}

\begin{figure}[h]
\centering
\begin{tabular}{cc}
\includegraphics[width=10cm]{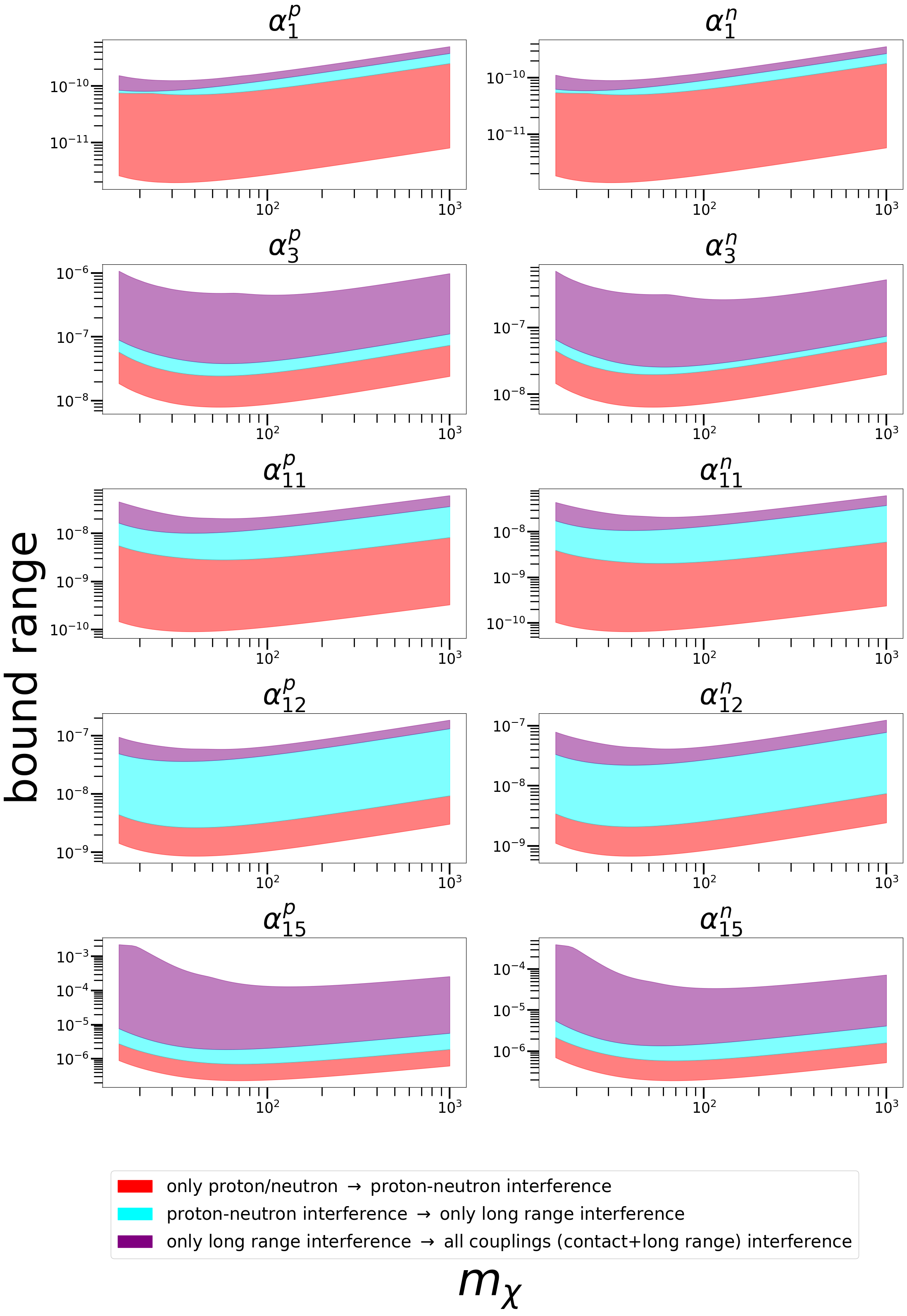} 
\end{tabular}
\caption{The same as in Fig.~\ref{fig:SI_shade} for the "spin--independent" long range WIMP--proton couplings $\alpha^p_{1}$, $\alpha^p_{3}$, $\alpha^p_{11}$, $\alpha^p_{12}$ and $\alpha^p_{15}$ (left column) and the WIMP--neutron couplings $\alpha^n_{1}$, $\alpha^n_{3}$, $\alpha^n_{11}$, $\alpha^n_{12}$ and $\alpha^n_{15}$ (left column). \label{fig:SI_long_shade}}
\end{figure}

\begin{figure}[h]
\centering
\begin{tabular}{cc}
\includegraphics[width=10cm]{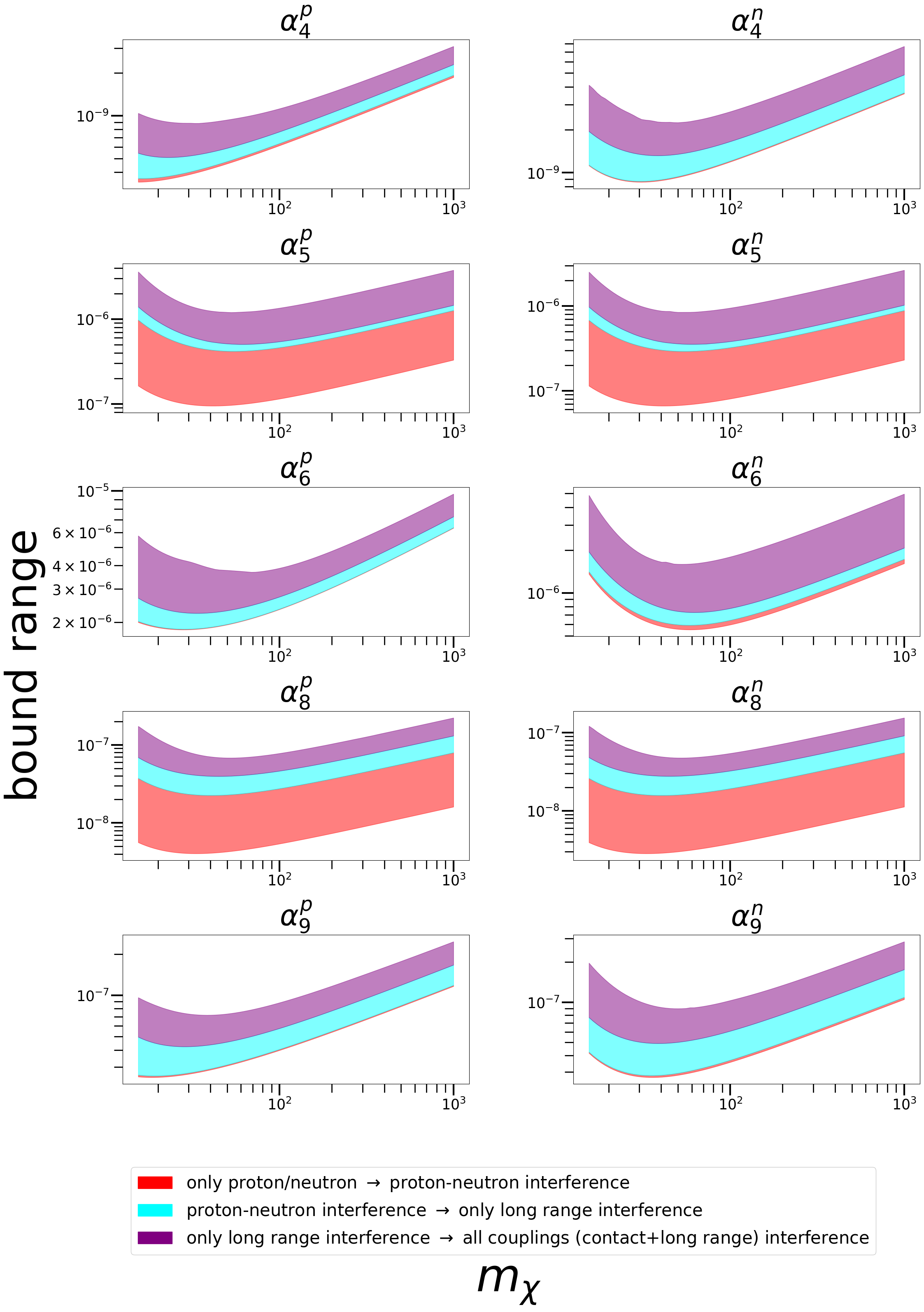} 
\end{tabular}
\caption{The same as in Fig.~\ref{fig:SI_shade} for the "spin--dependent" long range WIMP--proton couplings $\alpha^p_{4}$, $\alpha^p_{5}$, $\alpha^p_{6}$, $\alpha^p_{8}$, $\alpha^p_{9}$ (left column) and the WIMP--neutron couplings $\alpha^n_{4}$, $\alpha^n_{5}$, $\alpha^n_{6}$, $\alpha^n_{8}$, $\alpha^n_{9}$ (right column). \label{fig:SD_long_shade}}
\end{figure}

\begin{figure}[h]
\centering
\begin{tabular}{cc}
\includegraphics[width=10cm]{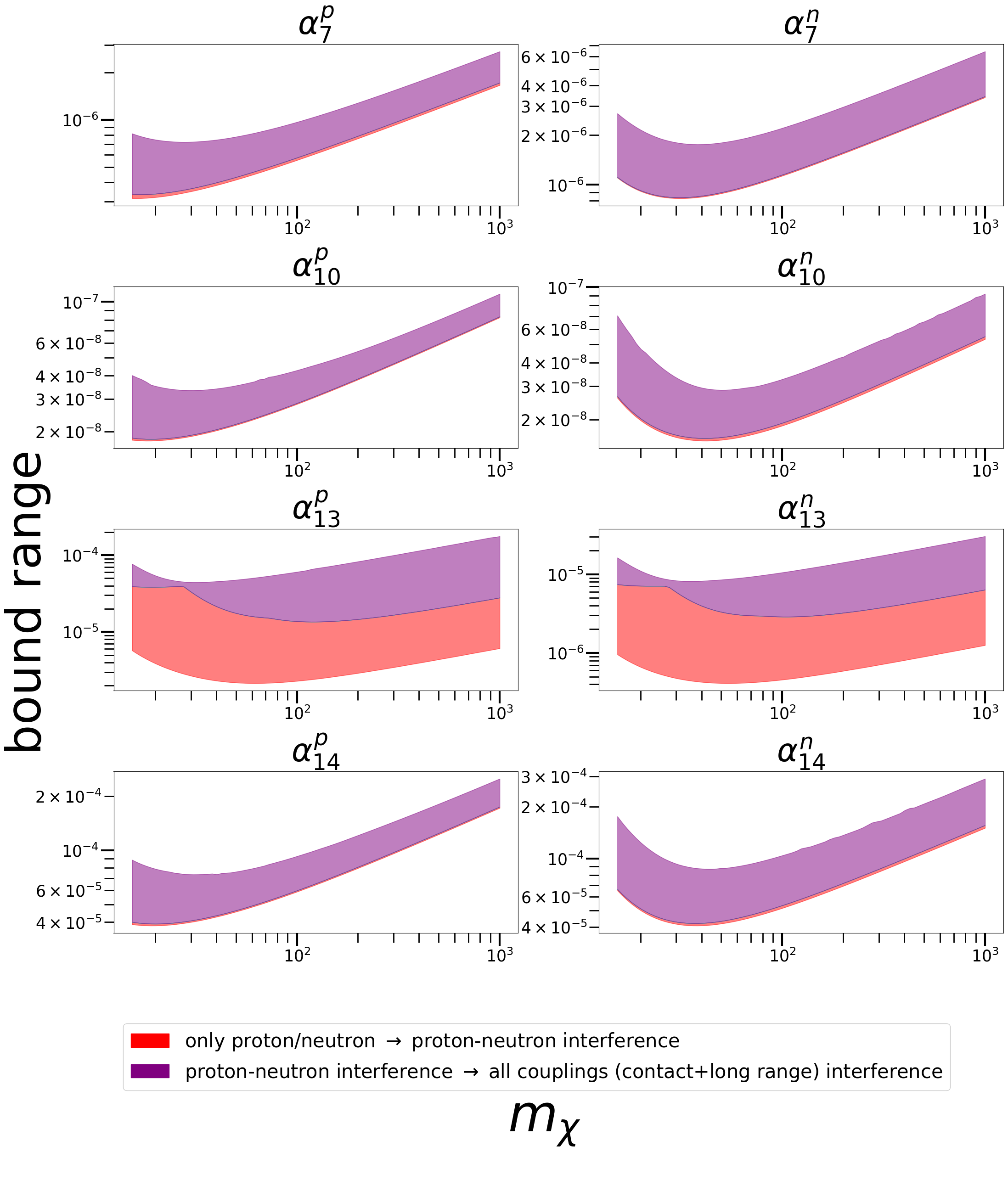} 
\end{tabular}
\caption{The same as in Fig.~\ref{fig:SI_shade} for the non-interfering long range WIMP--proton couplings $\alpha^p_{7}$, $\alpha^p_{10}$, $\alpha^p_{13}$, $\alpha^p_{14}$ (left column) and the WIMP--neutron couplings $\alpha^n_{7}$, $\alpha^n_{10}$, $\alpha^n_{13}$, $\alpha^n_{14}$ (right column) \label{fig:no_op_interf_long_shade}}
\end{figure}

The maximal variation of the exclusion plot (exclusion band) for each of the 56 couplings of the effective Hamiltonian of Eq.~(\ref{eq:H_short_long}) is shown in Figs.~\ref{fig:SI_shade}--\ref{fig:no_op_interf_long_shade}. In each plot the exclusion band is plotted as a function of the WIMP mass $m_\chi$ when the corresponding operator is allowed to interfere with a growing number of other interactions.  In particular in all the plots (i) the lower (red) band is spanned when each operator interferes in its own proton--neutron parameter space; (ii) the middle (cyan) band (if present) represents how the variation of the exclusion plot is extended when each operator $c^{p(n)}_k$ interferes also with the other couplings $c^{p(n)}_{k^{\prime}}$, or each $\alpha^{p(n)}_k$ interferes with the other couplings  $\alpha^{p(n)}_{k^{\prime}}$, with $k^{\prime}\ne k$ (contact/long range interactions); (iii) the upper (purple) band is the additional extension of the exclusion band when interferences between contact and long range interactions is also allowed (contact+long range interactions). All the configurations plotted in the figures are tested for convergence of the numerical optimization procedure described in Section~\ref{sec:conservative}.
In particular in all the plots at low WIMP masses such convergence is systematically not achieved. For this reason in all of them we put a cut at $m_\chi$ = 15 GeV. We explain this with the fact that when $m_\chi$ is small all signals are suppressed by the tail of the velocity distribution and their sensitivity to the parameters is enhanced. 

In our numerical solution of Eqs.~(\ref{eq:csi_lmi},\ref{eq:lmi}) we have included the nine matrices $A_k$ corresponding to the bounds from LZ~\cite{LZ_2022}, PandaX--4T~\cite{pandax4T_2021}, XENON1T~\cite{xenon_2018}, PICO--60($C_3F_8$)~\cite{pico60_2019}, PICO--60 ($CF_3I$)~\cite{pico60_2015}, SuperCDMS~\cite{super_cdms_2017}, CDMSlite~\cite{cdmslite_2017}, COSINE--100~\cite{cosine_bck} and DAMIC~\cite{DAMIC_2016} (see Appendix~\ref{app:experiments} for the details of their implementation). The optimization procedure to get the conservative upper bound $\max(c_\alpha)$ on the coupling $c_\alpha$ yields an extreme couplings vector $\bm{c}$ with $\max(|c_\alpha|)$ = $|\bm{c}_\alpha|$. Experiments for which $\bm{c}^T\cdot A_k\cdot\bm{c}$ = 1 saturate their own bound and contribute to reducing the allowed parameter space, while those for which $\bm{c}^T\cdot A_k\cdot\bm{c}<$ 1 are not sensitive enough and do not play any role in determining $\max(|c_\alpha|)$. The general result of our analysis is that for all the 56 couplings analyzed only LZ~\cite{LZ_2022}, PandaX--4T~\cite{pandax4T_2021}, PICO--60($C_3F_8$) and PICO--60 ($CF_3I$) saturate their own bounds and contribute to determine $\max(|c_\alpha|)$.       

From the expressions of the WIMP response functions of Eq.~(\ref{eq:wimp_response_functions}) one can see that not all the operators in Eq.~(\ref{eq:H_short_long}) interfere, but, instead, the 56--dimensional parameter space breaks down in the following non--interfering subspaces:  $[1,3]$, $[8,9]$, $[4,5,6]$ and $[11,12,15]$. Moreover in the subspaces  $[7]$, $[10]$, $[13]$ and $[14]$ there is no interference among different couplings, but only within the proton--neutron and/or contact/long range component of each coupling. As a consequence, in order to find the conservative bounds it is sufficient to solve Eqs.~(\ref{eq:csi_lmi}, \ref{eq:lmi}) in the non--interfering subspaces rather than in the full parameter space and this greatly simplifies the task numerically. In particular at most the dimensionality of each subspace in our analysis can reach 12 (in the case of 3 interfering operators ${\cal O}_k$ with the two proton and neutron components and for short+long interactions).

Moreover, (see Eq.~(\ref{eq:r_decomposition}) and Table~\ref{table:eft_summary}) according to the type of nuclear form factor that drives their interaction, both contact and long range couplings can be classified in different groups~\cite{nreft_sensitivity_sogang}.

A first class of "spin--dependent"--type interactions corresponds to the effective operators ${\cal O}_4$, ${\cal O}_6$, ${\cal O}_7$, ${\cal O}_9$, ${\cal O}_{10}$ and ${\cal O}_{14}$, that are driven by either the nuclear form factor $W^{\tau\tau^{\prime}}_{\Sigma^{\prime\prime}}$ or by $W^{\tau\tau^{\prime}}_{\Sigma^{\prime}}$, which couple the WIMP to the nuclear spin (the sum $W^{\tau\tau^{\prime}}_{\Sigma^{\prime\prime}}+W^{\tau\tau^{\prime}}_{\Sigma^{\prime}}$ corresponds to the standard spin--dependent form factor~\cite{menendez}). 
The exclusion bands of "spin--dependent" operators that interfere with other ones, as well as those couplings that belong to the same interfering subspaces (i.e, $[4,5,6]$ and $[8,9]$), are shown in Figs.~\ref{fig:SD_shade} and \ref{fig:SD_long_shade}. On the other hand the "spin--dependent" couplings driven by ${\cal O}_7$, ${\cal O}_{10}$ and ${\cal O}_{14}$ are shown in Figs.~\ref{fig:no_op_interf_shade} and \ref{fig:no_op_interf_long_shade}, where we have grouped those operators that do not interfere with others.

Since inside nuclei the nucleons spins tend to cancel each other the contribution from even--numbered nucleons to the response functions $\Sigma^{\prime\prime}$ and $\Sigma^{\prime}$ is
strongly suppressed. As a consequence, for such interactions the flat directions discussed in Section~\ref{sec:conservative} are mostly aligned to the proton axes for neutron--odd targets (such as xenon and germanium), while they point along the neutron axes for proton--odd targets
(such as fluorine and iodine)\cite{complementarity_munich_sogang}. This means that for this class of interactions combining proton--odd and neutron--odd targets according to the procedure discussed in Section~\ref{sec:combining} is particularly effective in constraining the allowed parameter space. This explains why the plots of Figs.~~\ref{fig:SD_shade} and \ref{fig:SD_long_shade} (for interfering couplings) and those in Figs.~\ref{fig:no_op_interf_shade} and \ref{fig:no_op_interf_long_shade} (for the non--interfering ones) show a moderate extension of the exclusion band, especially in the proton--neutron interference subspace (lower red band). Indeed, in some cases {\it the width of the most conservative exclusion band does not exceed a factor of a few} (${\cal O}_4$, ${\cal O}_6$, ${\cal O}_7$, ${\cal O}_9$, ${\cal O}_{10}$, ${\cal O}_{14}$). For this class of interactions we find that only  the combination of PandaX--4T~\cite{pandax4T_2021} and/or LZ~\cite{LZ_2022}, (that use $Xe$, a neutron--odd target), PICO--60 ($C_3F_8$) and PICO--60 ($CF_3I$) (proton--odd targets) can contribute to the bounds, while the other experiments turn out to be irrelevant in constraining the parameter space.

A second class of "spin--independent"--type interactions consists in the operators ${\cal O}_1$, ${\cal O}_3$, ${\cal O}_{11}$, ${\cal O}_{12}$ and ${\cal O}_{15}$, that are all driven either by $W^{\tau\tau^{\prime}}_{M}$ or $W^{\tau\tau^{\prime}}_{\Phi^{\prime\prime}}$. Such interactions are both enhanced for heavy targets. $M$ corresponds to the standard Spin Independent coupling, proportional to the nuclear mass number squared; on the other hand, $\Phi^{\prime\prime}$ is non vanishing for all nuclei and favors heavier elements with large nuclear shell model orbitals not fully occupied. Its scaling with the nuclear target is similar to the SI interaction, albeit the corresponding nuclear response functions are about two orders of magnitude smaller. As a consequence, both $M$ and $\Phi^{\prime\prime}$  rather than showing a complementarity among different targets tend to favour xenon in PandaX--4T~\cite{pandax4T_2021} and  LZ~\cite{LZ_2022}, that can drive the limit alone also thanks to their large exposures, although enlarging the couplings subspace also PICO--60($C_3F_8$) and PICO--60($CF_3I$) can contribute to the bound. This class of "spin--independent" interactions is shown in Figs.~\ref{fig:SI_shade} and \ref{fig:SI_long_shade}, from which one can see that due to the reduced degree of complementarity among different targets the exclusion bands can span a wider range compared to "spin--dependent" ones, reaching a variation of up to three orders of magnitude.  

As already pointed out, in Figs.~\ref{fig:no_op_interf_shade} and \ref{fig:no_op_interf_long_shade} we have grouped those operators that do not interfere with others, for a contact interaction and for a long range one, respectively. In this case the possible interferences 
are between the proton and neutron coupling ($[c^p_k, c^n_k]$ or $[\alpha^p_k, \alpha^n_k]$ subspaces) or include the interference between short and long range interactions ($[c^p_k,c^n_k, \alpha^p_k,\alpha^n_k]$ subspace). As explained above the plots in Figs.~\ref{fig:no_op_interf_shade} and \ref{fig:no_op_interf_long_shade} 
driven by ${\cal O}_7$, ${\cal O}_{10}$ and ${\cal O}_{14}$ are of the "spin--dependent" type, and have a moderate extension of the exclusion bands. Moreover Figs.~\ref{fig:no_op_interf_shade} and \ref{fig:no_op_interf_long_shade} include also the interactions driven by ${\cal O}_{13}$ ($c_{13}^{(p,n)}$ and $\alpha_{13}^{(p,n)}$), which is the only operator for a spin 1/2 particle driven by $W^{\tau\tau^{\prime}}_{\tilde{\Phi}^{\prime}}$. Such nuclear form factor requires a target spin $j_T> 1/2$, a property of only $^{23}Na$, $^{73}Ge$, $^{127}I$ and $^{131}Xe$ among the isotopes used in DM searches. We observe that the corresponding bounds are either driven by PandaX--4T~\cite{pandax4T_2021} or LZ~\cite{LZ_2022} alone or determined by a combination of xenon targets and scattering off fluorine in PICO--60($C_3F_8$). In such case, since the spins for fluorine and carbon are $1/2$ and $0$, respectively, the cross section in PICO--60 takes only contribution from the velocity--dependent part (i.e. from $R^{\tau\tau^{\prime}}_{1\tilde{\Phi}^{\prime}}$ in Eq.~(\ref{eq:r_decomposition})) off fluorine, which, as shown in Table~\ref{table:eft_summary} depends on the $\Sigma^{\prime\prime}$.
The peculiarity that for ${\cal O}_{13}$, in spite of the suppression due to the WIMP speed, the velocity--dependent part of the rate off fluorine in  PICO--60  can be as constraining as the velocity–independent rate off xenon was already pointed out in the analysis of Ref.~\cite{nreft_sensitivity_sogang} for XENON1T.

\begin{table}[t]
\begin{center}
{\begin{tabular}{|@{}|c|c|c|c|c|c|@{}}
%{@{}|c|c|c|c|c|c|@{}}
\hline
coupling  &  $R^{\tau \tau^{\prime}}_{0k}$  & $R^{\tau \tau^{\prime}}_{1k}$ & coupling  &  $R^{\tau \tau^{\prime}}_{0k}$  & $R^{\tau \tau^{\prime}}_{1k}$ \\
\hline
$1$  &   $M(q^0)$ & - & $3$  &   $\Phi^{\prime\prime}(q^4)$  & $\Sigma^{\prime}(q^2)$\\
$4$  & $\Sigma^{\prime\prime}(q^0)$,$\Sigma^{\prime}(q^0)$   & - & $5$  &   $\Delta(q^4)$  & $M(q^2)$\\
$6$  & $\Sigma^{\prime\prime}(q^4)$ & - & $7$  &  -  & $\Sigma^{\prime}(q^0)$\\
$8$  & $\Delta(q^2)$ & $M(q^0)$ & $9$  &  $\Sigma^{\prime}(q^2)$  & - \\
$10$  & $\Sigma^{\prime\prime}(q^2)$ & - & $11$  &  $M(q^2)$  & - \\
$12$  & $\Phi^{\prime\prime}(q^2)$,$\tilde{\Phi}^{\prime}(q^2)$ & $\Sigma^{\prime\prime}(q^0)$,$\Sigma^{\prime}(q^0)$ & $13$  & $\tilde{\Phi}^{\prime}(q^4)$  & $\Sigma^{\prime\prime}(q^2)$ \\
$14$  & - & $\Sigma^{\prime}(q^2)$ & $15$  & $\Phi^{\prime\prime}(q^6)$  & $\Sigma^{\prime}(q^4)$ \\
\hline
\end{tabular}}
\caption{Nuclear response functions corresponding to each coupling, for the velocity--independent and the velocity--dependent components parts of the WIMP response function, decomposed as in Eq.(\ref{eq:r_decomposition}).
 In parenthesis the power of $q$ in the WIMP response function is shown.
  \label{table:eft_summary}}
\end{center}
\end{table}

\begin{figure}[h]
\centering
\begin{tabular}{cc}
\includegraphics[width=6cm]{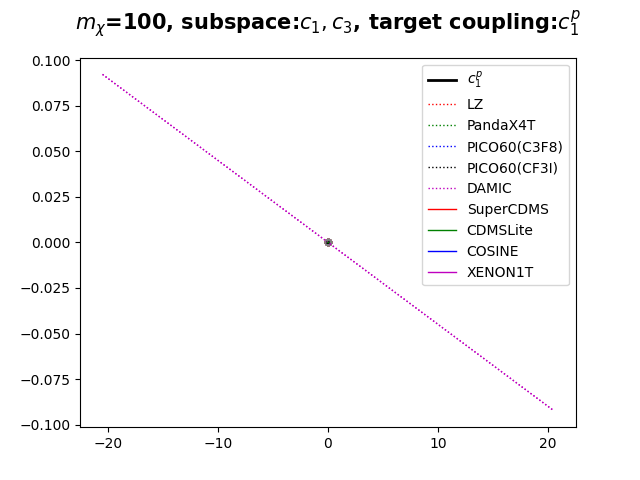} 
\includegraphics[width=6cm]{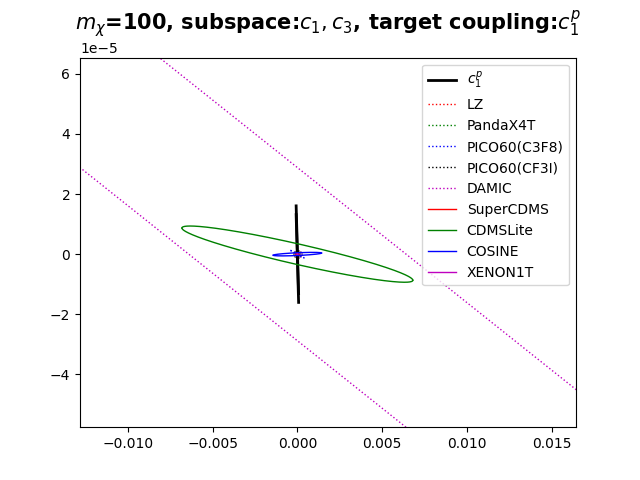} \\
\includegraphics[width=6cm]{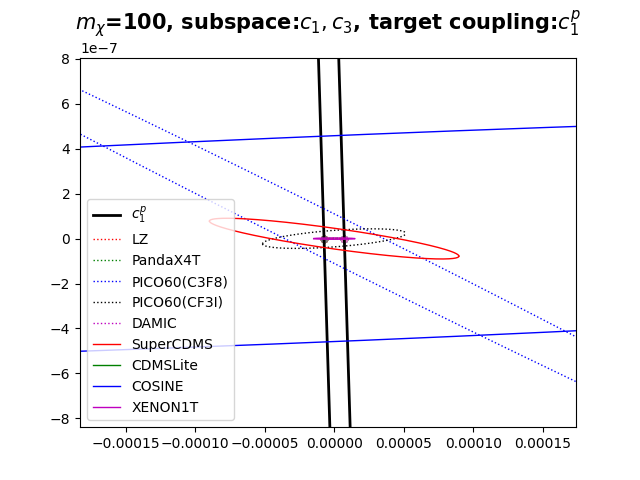} 
\includegraphics[width=6cm]{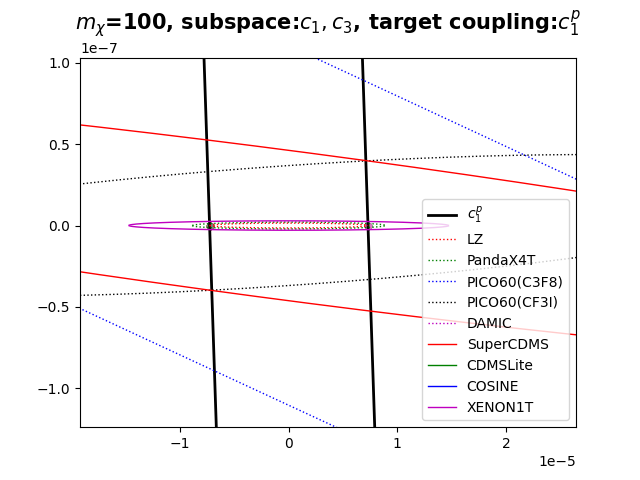} \\
\includegraphics[width=6cm]{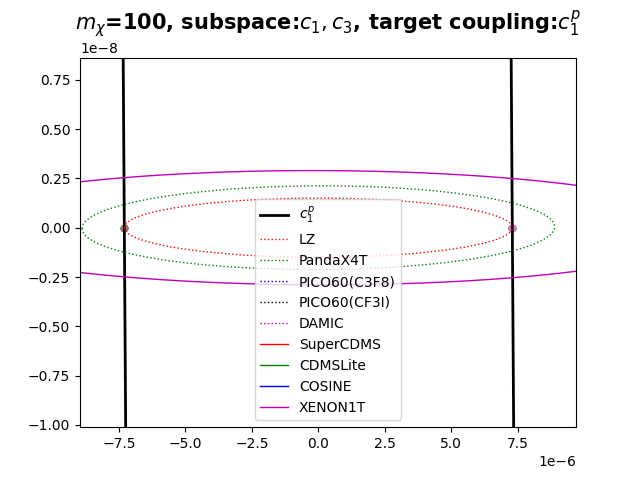}
\includegraphics[width=5cm]{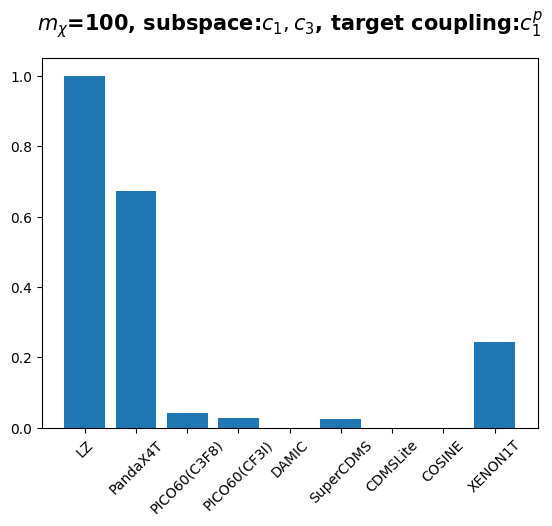}
\end{tabular}
\caption{Different magnifications of the two--dimensional projections of the ellipsoids that determine the upper bound on $c^p_1$ for $m_\chi$ = 100 GeV when interferences with other contact interactions ($c^n_1$, $c^p_3$, $c^n_3$) are included. The projection plane is rotated so that the optimized couplings vector $\bm{c}$ lies on the horizontal axes. The two parallel solid lines represent the intersection of the two hyperplanes $(c^p_1)^2\le max(|c^p_1|)^2$ with the projection plane and nicely cross the points that delimit the allowed parameter region, confirming a successful convergence. The last plot contains the corresponding $\bm{c}^T\cdot A_k \cdot \bm{c}$ values for the nine experiments included in the analysis (see Eq.~(\ref{eq:lmi})).  \label{fig:c1_ellipsoids}}
\end{figure}

\begin{figure}[h]
\centering
\begin{tabular}{cc}
\includegraphics[width=7cm]{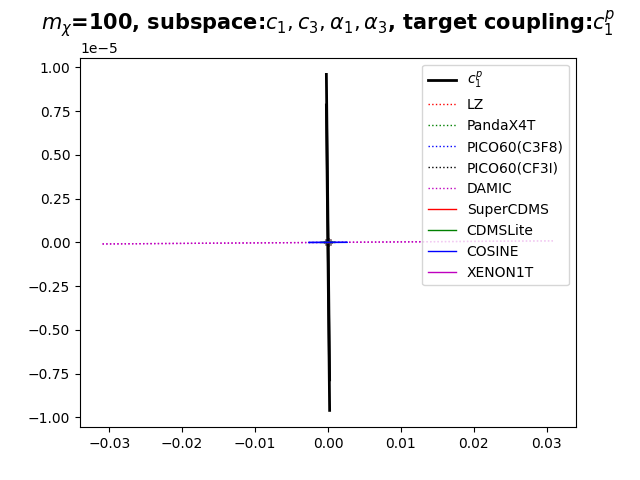} 
\includegraphics[width=7cm]{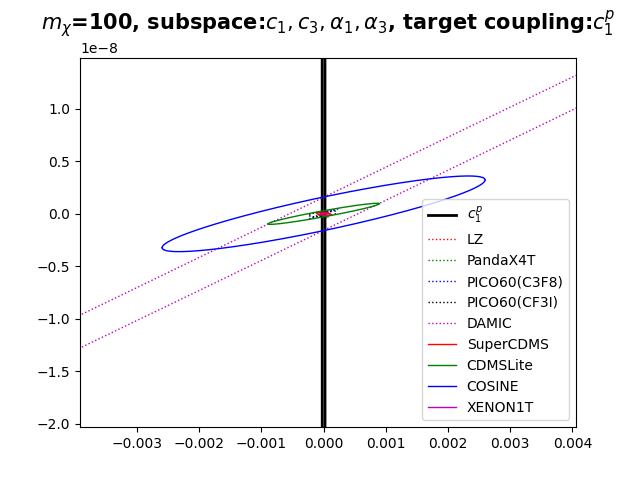} \\
\includegraphics[width=7cm]{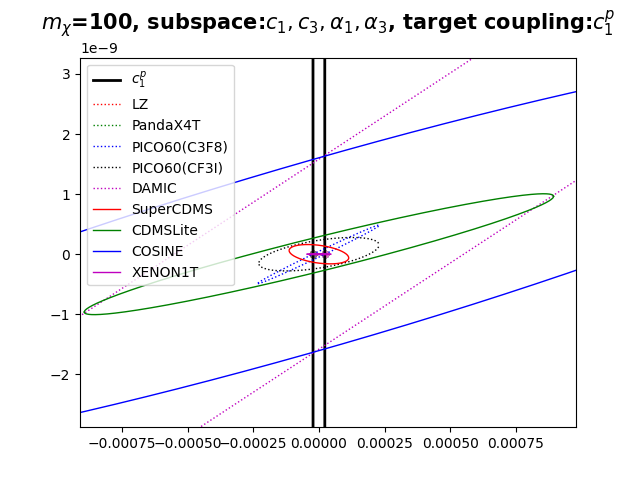} 
\includegraphics[width=7cm]{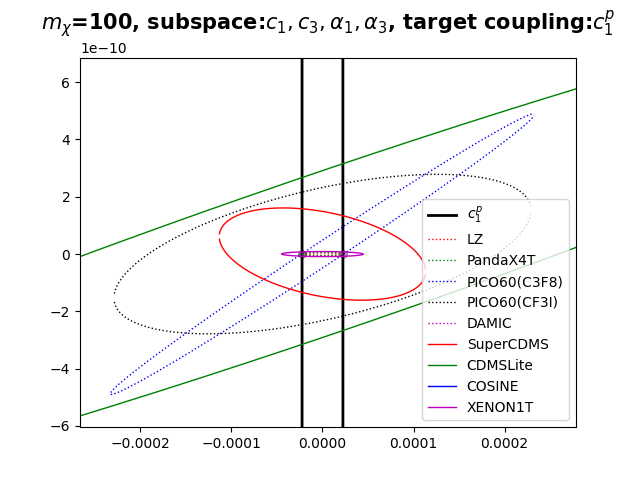} \\
\includegraphics[width=7cm]{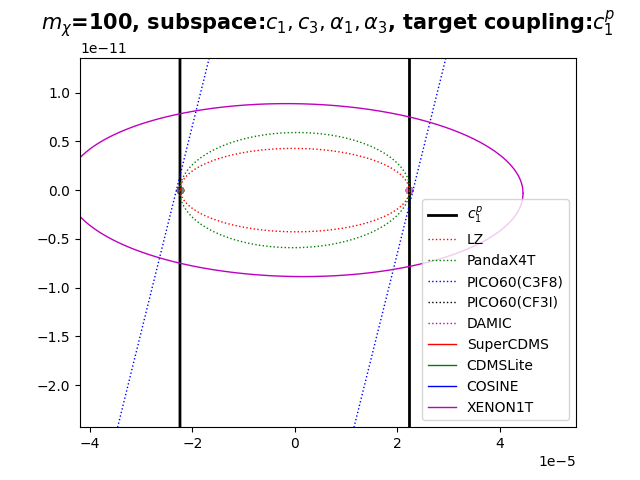}
\includegraphics[width=6cm]{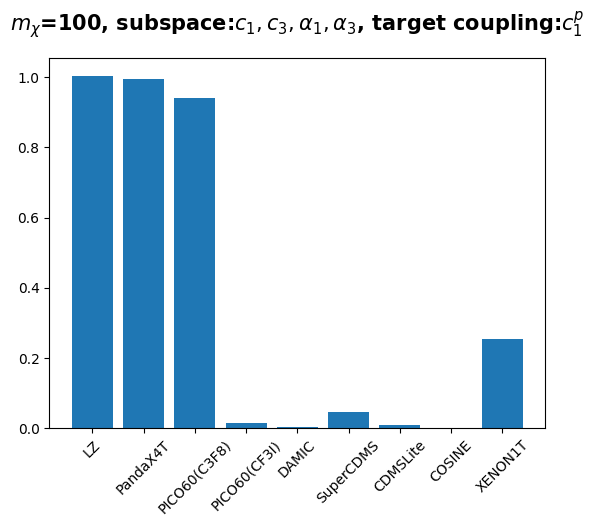}
\end{tabular}
\caption{The same of Fig.~\ref{fig:c1_ellipsoids} when interferences with both other contact ($c^n_1$, $c^p_3$, $c^n_3$) and long range ($\alpha^p_1$, $\alpha^n_1$, $\alpha^p_3$, $\alpha^n_3$) interactions are included. Also in this case 
the two parallel solid lines that represent the intersection of the two hyperplanes $(c^p_1)^2\le max(|c^p_1|)^2$ with the projection plane nicely cross the points that delimit the allowed parameter region, given by the intersection of the LZ and PandaX--4T ellipsoids, confirming a successful convergence. \label{fig:c1_full_ellipsoids}}
\end{figure}

\begin{figure}[h]
\centering
\begin{tabular}{cc}
\includegraphics[width=7cm]{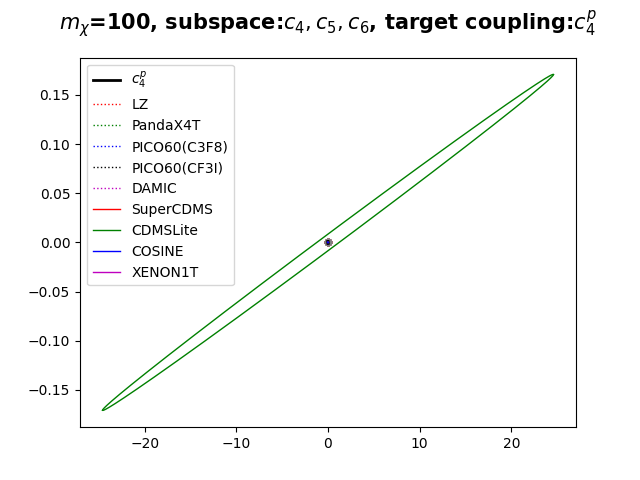} 
\includegraphics[width=7cm]{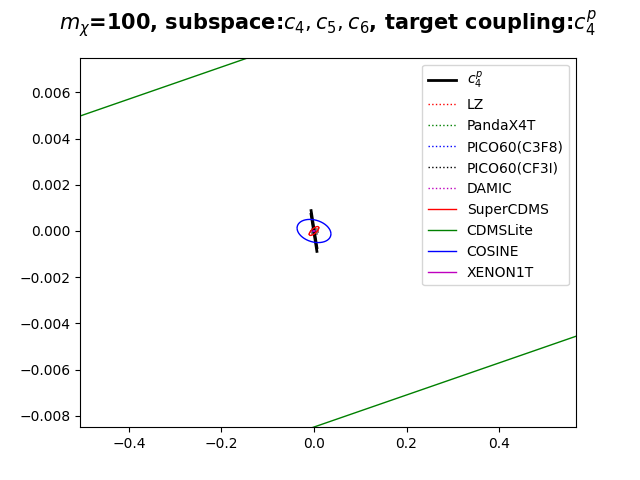} \\
\includegraphics[width=7cm]{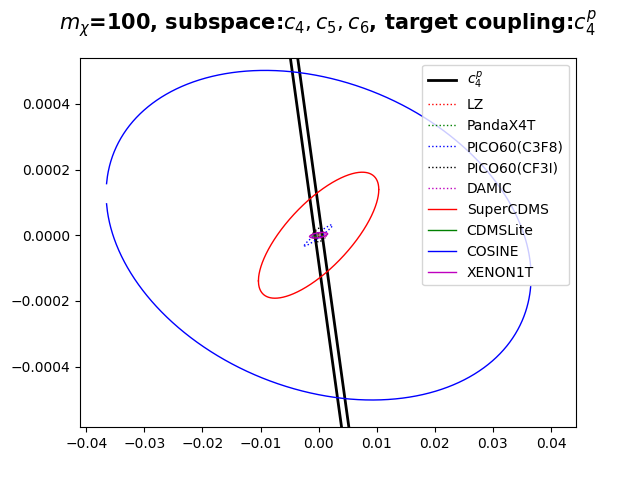}
\includegraphics[width=7cm]{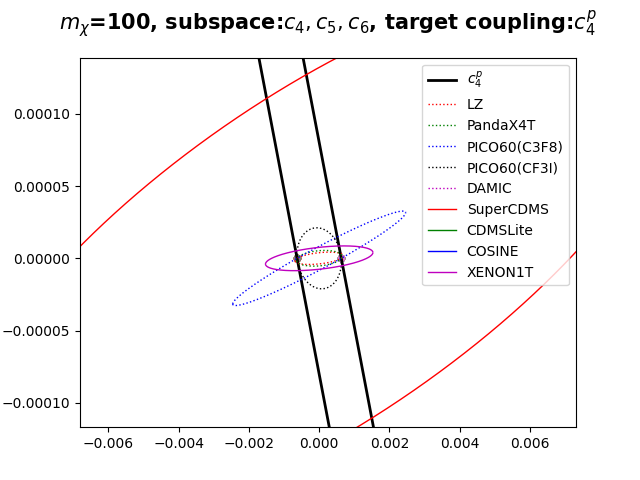} \\
\includegraphics[width=7cm]{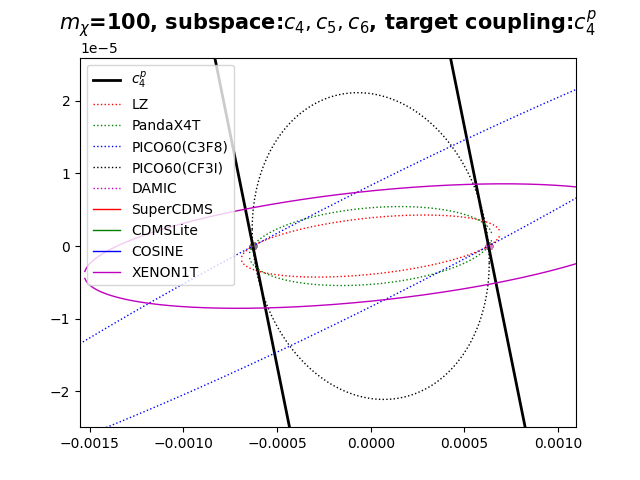}
\includegraphics[width=6cm]{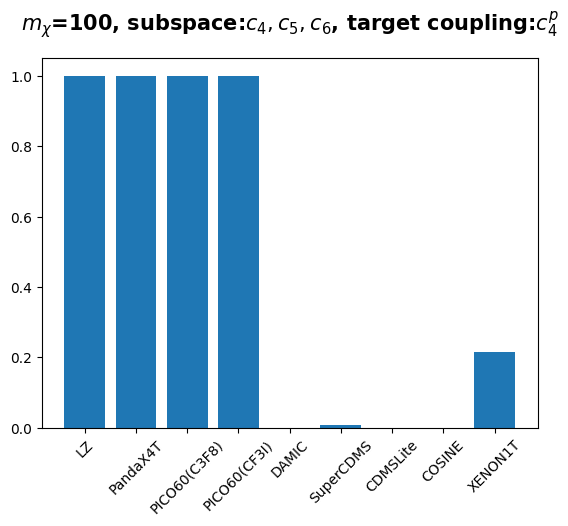}
\end{tabular}
\caption{The same of Fig.~\ref{fig:c1_ellipsoids} for the upper bound on $c^p_4$ when interferences with other short--range interactions ($c^n_4$, $c^p_5$, $c^n_5$, $c^p_6$, $c^n_6$) are included. In this case four experiments (LZ, PandaX--4T, PICO--60($C_3F_8$) and PICO--60 ($CF_3I$)) saturate their bounds and determine the constraint on $c^p_4$.\label{fig:c4_ellipsoids}}
\end{figure}

In Figs.~\ref{fig:c1_ellipsoids}, \ref{fig:c1_full_ellipsoids} and \ref{fig:c4_ellipsoids} we provide plots of our optimization procedure for some specific examples. In Fig.~\ref{fig:c1_ellipsoids} the nine ellipsoids corresponding to the conservative upper bound on $c^p_1$ are projected in a two--dimensional plane, and a histogram of the corresponding $\bm{c}^T\cdot A_k \cdot \bm{c}$ is provided. The projection plane is rotated so that the optimized couplings vector $\bm{c}$ lies on the horizontal axis. Different magnifications of the same plot are provided to make all the ellipsoids visible. In this particular example only LZ  determines the bound and saturates its own constraint. This is also shown in the histogram where only the $\bm{c}^T\cdot A_k \cdot \bm{c}$ value for LZ is equal to one.
In this case LZ alone determines the bound because $c^p_1$ is a spin--independent coupling that favours heavy nuclei. 

In Fig.~\ref{fig:c1_full_ellipsoids} the nine ellipsoids corresponding to the conservative upper bound on $c^p_1$ are plotted when interferences with both other contact and long range interactions are included. In this case the two xenon detectors LZ and PandaX--4T drive the constraint on the parameter space.

A final example is provided in Fig.~\ref{fig:c4_ellipsoids} for the case of $c^p_4$ when only interferences with other contact operators are considered. Here LZ, PandaX--4T, PICO--60($C_3F_8$) and PICO--60($CF_3I$) determine the bound.  

Notice how in Figs.~\ref{fig:c1_ellipsoids}, \ref{fig:c1_full_ellipsoids} and \ref{fig:c4_ellipsoids} the two parallel solid lines representing the intersection of the hyperplanes that encompass the maximal excursion of the target coupling with the projection plane nicely cross the points that delimit the allowed parameter region, confirming a successful convergence.

\begin{figure}[h]
\centering
\begin{tabular}{cc}
\includegraphics[width=7cm]{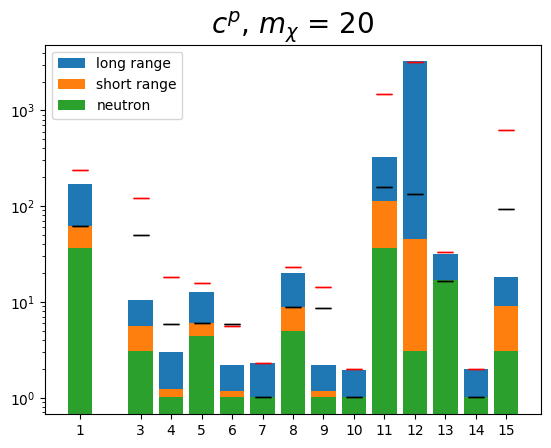} 
\includegraphics[width=7cm]{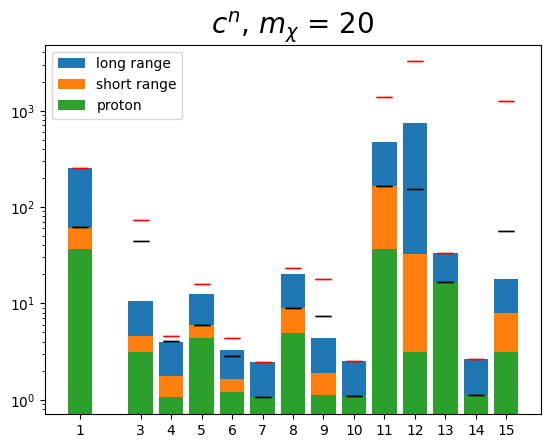} \\
\includegraphics[width=7cm]{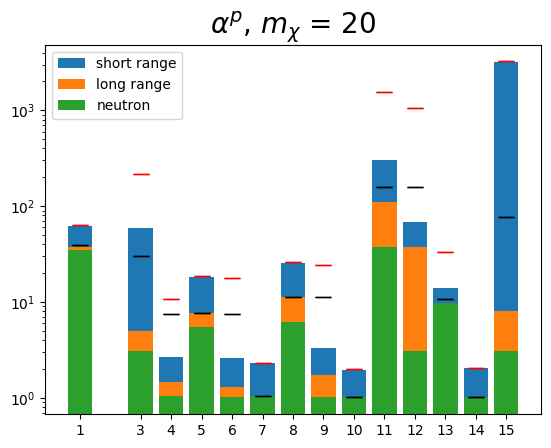}
\includegraphics[width=7cm]{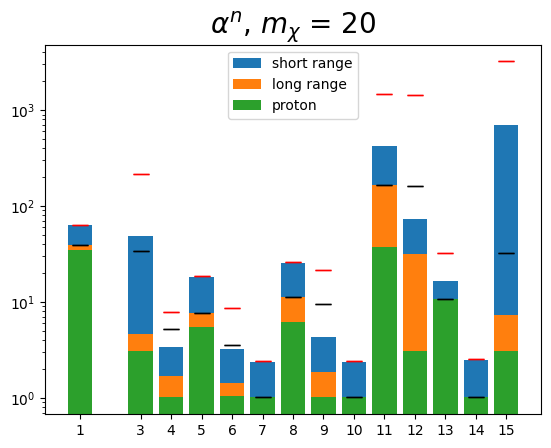}
\end{tabular}
\caption{Relaxation factors between the less constraining and the most constraining bounds for $m_{\chi}$ = 20 GeV. Upper row: contact interactions; lower row: long range interactions. Left column: WIMP--proton couplings; right column: WIMP--neutron couplings. 
%The color code is the same of the  exclusion bands plots discussed in Figs.~\ref{fig:SI_shade}, \ref{fig:SD_shade}, \ref{fig:no_op_interf_shade}, \ref{fig:SI_long_shade}, \ref{fig:SD_long_shade} and \ref{fig:no_op_interf_long_shade}.
Horizontal bars: square--root of the tuning factor $\xi^{1/2}$ (see Eqs.~(\ref{eq:tuning}) and (\ref{eq:xi})) 
 for interferences among only contact or only long range couplings (black) or for interferences including both contact and long range couplings (red).
\label{fig:relaxing_factor_20}}
\end{figure}

\begin{figure}[h]
\centering
\begin{tabular}{cc}
\includegraphics[width=7cm]{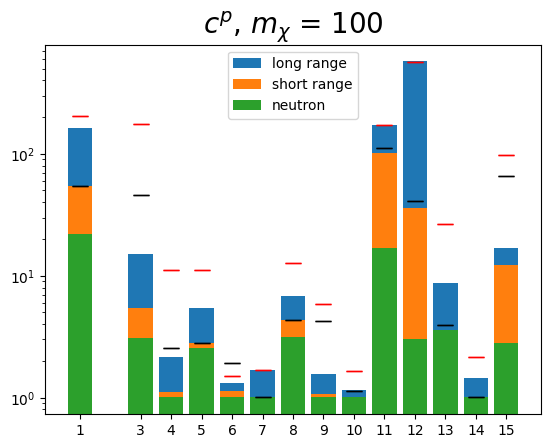} 
\includegraphics[width=7cm]{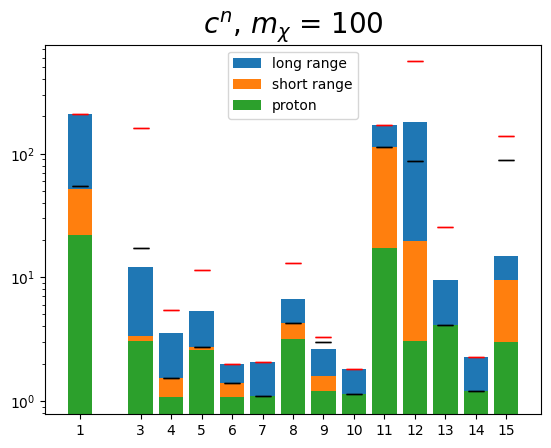} \\
\includegraphics[width=7cm]{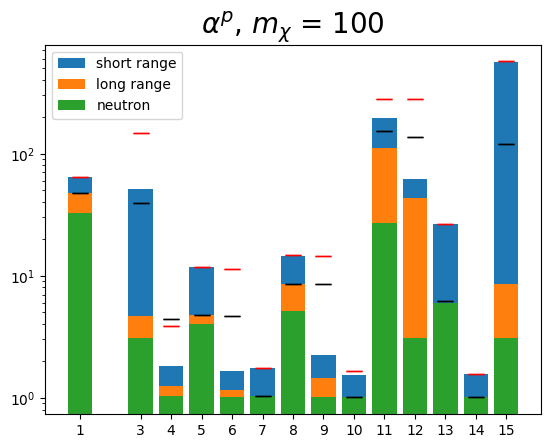}
\includegraphics[width=7cm]{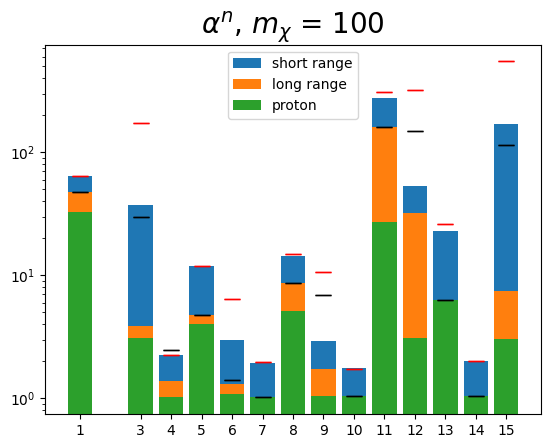}
\end{tabular}
\caption{Same as fig~\ref{fig:relaxing_factor_20} for $m_{\chi}$ = 100 GeV. \label{fig:relaxing_factor_100} }
\end{figure}

\begin{figure}[h]
\centering
\begin{tabular}{cc}
\includegraphics[width=7cm]{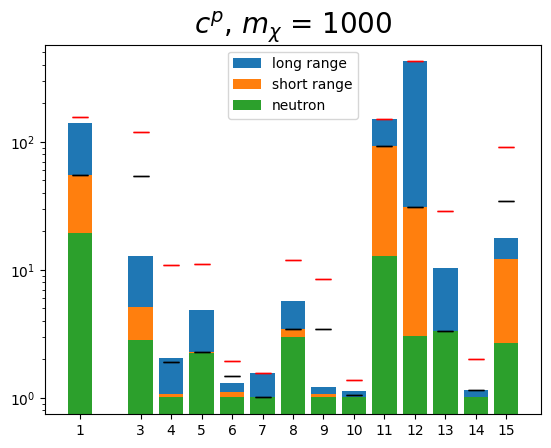} 
\includegraphics[width=7cm]{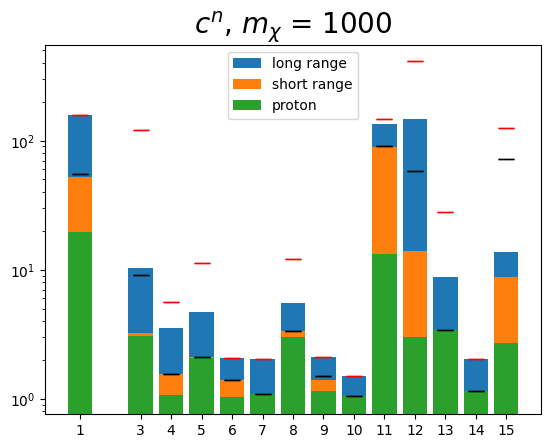} \\
\includegraphics[width=7cm]{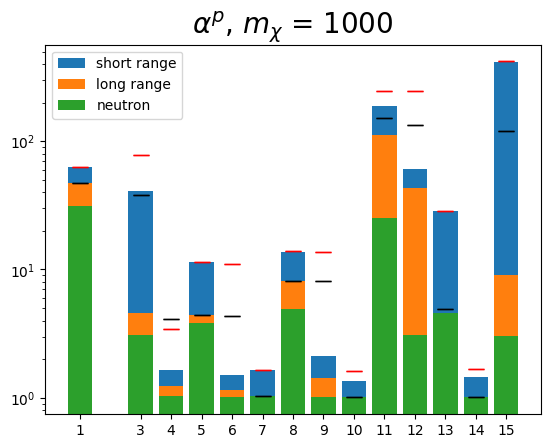}
\includegraphics[width=7cm]{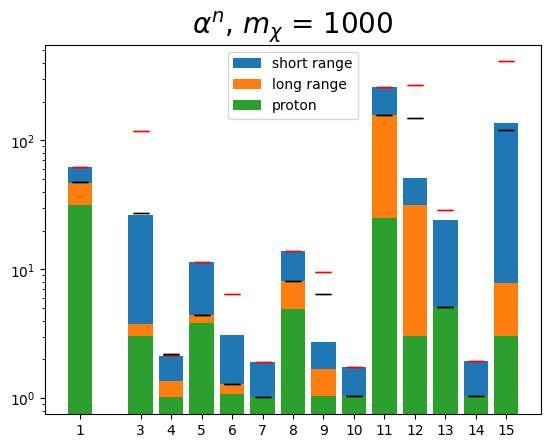}
\end{tabular}
\caption{Same as fig~\ref{fig:relaxing_factor_20} for $m_{\chi}$ = 1000 GeV. \label{fig:relaxing_factor_1000}}
\end{figure}

\subsection{Relaxation factors and sensitivity of the results}
\label{sec:cancellations}

The histograms of Figs.~\ref{fig:relaxing_factor_20}, \ref{fig:relaxing_factor_100} and \ref{fig:relaxing_factor_1000} show for $m_\chi$=20 GeV, 100 GeV and 1 TeV, respectively, the relaxation factor $r$ = $c_k^{less}/c_k^{most}$ ($\alpha_k^{less}/\alpha_k^{most}$) between the less constraining and the most constraining bounds $c_k^{less}$ and $c_k^{most}$ ($\alpha_k^{less}$ and $\alpha_k^{most}$) on each of the couplings $c_k^{(p,n)}$ and $\alpha_k^{(p,n)}$, under the same conditions of the  exclusion bands  discussed in the previous Section. Such plots allow to directly compare the width of the exclusion bands of the different couplings. 

In particular from such figures one can observe that for some operators the relaxation factor can be sizeable, exceeding in few cases three orders of magnitude. In such situations not only the exclusion bands of Figs.~\ref{fig:SI_shade}--\ref{fig:no_op_interf_long_shade} turn out to be so large to be of little practical use: more importantly their reliability is put into question, since  to obtain such a large relaxation of the bounds a high level of cancellation in the calculation of the expected rate is expected, and this should be related to a high sensitivity of the result on the values of the input matrices, as discussed in~\cite{complementarity_munich_sogang}.
In presence of such a large sensitivity to obtain
a robust conservative upper bound it would be in principle necessary to include higher--order effects in the calculation of the scattering rates, in particular the sub--dominant contributions from two--nucleon scattering~\cite{Cirigliano_two_bodies_2012, Klos_two_bodies_2013, Vietze_two_bodies_2014, Cirigliano_two_bodies_2014}. As already pointed out combining different experiments can solve, or alleviate, this issue. A qualitative graphic explanation of the reason is illustrated in Fig.~\ref{fig:ellipses_bracketing}, where the combined bound $\max(c_\alpha)_{12}$ is not affected when the ellipse for Exp 1 is changed to Exp 1$^{\prime}$ due to a near--to--vanishing eigenvalue that makes the corresponding matrix very sensitive to small changes of its numerical entries. However, besides qualitative explanations, it is crucial to understand in a quantitative way to which extent this is achieved. 

The most constraining bound on a given coupling $c_\alpha$ is obtained by assuming that it is the only non--vanishing one. For a given experiment and energy bin with matrix ${\cal M}$, assuming a contact interaction:

\begin{equation}
 c^{most}_\alpha {\cal M}_{\alpha\alpha}c^{most}_\alpha=\frac{1}{r^2}  c^{less}_\alpha {\cal M}_{\alpha\alpha}c^{less}_\alpha =1,
  \label{eq:c_most}
\end{equation}

\noindent while for the conservative bound the optimization problem of Eqs.~(\ref{eq:csi_lmi}, \ref{eq:lmi}) yields a couplings vector $\bm{c}^{less}$ with $c^{less}_\alpha$ = $[\bm{c}^{less}]_\alpha$, so that:
\begin{equation}
[\bm{c}^{less}]^T\cdot {\cal M} \cdot \bm{c}^{less}=\sum_{ij} [\bm{c}^{less}]_i {\cal M}_{ij} [\bm{c}^{less}]_j = r^2 + \sum_{ij\ne \alpha\alpha} [\bm{c}^{less}]_i {\cal M}_{ij} [\bm{c}^{less}]_j=1.
\label{eq:tuning}
\end{equation}

\noindent  Eq.~(\ref{eq:tuning}) shows that if $r\gg$ 1 
the signal prediction normalized to the bound can be equal to unity only through a large level of cancellation, at least equal to the square of the relaxation factor $r^2$, i.e. $r^2 + (-r^2+1)=1$.  A more accurate way to quantify the level of tuning $\xi$ is to identify it with the maximal contribution (in absolute value) to the sum that determines the signal prediction:

\begin{equation}
    \xi \equiv \max\left(\left |[\bm{c}^{less}]_i {\cal M}_{ij} [\bm{c}^{less}]_j\right| \right ) = \left|[\bm{c}^{less}]_\gamma {\cal M}_{\gamma\delta} [\bm{c}^{less}]_\delta\right| \ge r^2.
\label{eq:xi}
\end{equation}

\noindent The value of $\xi$ can be larger than $r^2$, with $\xi$ = $r^2$ only when $\gamma\delta$ = $\alpha\alpha$. In particular if the matrix element $\gamma\delta$ of ${\cal M}$ is modified as ${\cal M}_{\gamma\delta}\rightarrow {\cal M}_{\gamma\delta} (1+\epsilon)$ by a perturbation $\epsilon$,  by definition $[\bm{c}^{less}]^T\cdot {\cal M} \cdot \bm{c}^{less}$ is modified by $\pm$1 if $|\epsilon|$ = $1/\xi$, i.e. to have a reliable prediction the matrix element  ${\cal M}_{\gamma\delta}$ needs to be calculated with a precision better than $1/\xi$.

In Figs.~\ref{fig:relaxing_factor_20}, \ref{fig:relaxing_factor_100} and \ref{fig:relaxing_factor_1000} the values of $\xi^{1/2}$ are plotted as black (red) horizontal bars in the two cases of a short/long range and a short+long range interaction, respectively. Indeed, in several cases $\xi^{1/2}>r$. The same plots show that the broad classification made in Section~\ref{sec:exclusion_bands}  between "spin--dependent" type and "spin--independent" type interactions directly reflects in the tuning. 
In particular $\epsilon\gsim {\cal O}(10^{-3})$ for all the "spin--dependent" couplings (but $\epsilon\gsim {\cal O}(10^{-1})$ for ${\cal O}_{7}$, ${\cal O}_{10}$ and ${\cal O}_{14}$ and in most cases when only contact or long range interactions are included in the interferences). On the other hand the tuning for the couplings driven by "spin--independent" nuclear form factors is typically higher, with ${\cal O}(10^{-6})<\epsilon<{\cal O}(10^{-4})$  for ${\cal O}_{11}$, ${\cal O}_{12}$ and ${\cal O}_{15}$, ${\cal O}(10^{-5})<\epsilon<{\cal O}(10^{-3})$  for ${\cal O}_{1}$ and ${\cal O}_{3}$, and ${\cal O}(10^{-4})<\epsilon<{\cal O}(10^{-2})$  for ${\cal O}_{13}$.
For the evaluation of the matrices we use the routine \verb|wimp_dd_matrix|, released in a new version of \verb|WimPyDD|~\cite{wimpydd_2022}, that calculates expected rates by interpolating response functions tabulated as a function of the recoil energy. We have performed our calculation increasing the default sampling of the response functions from 100 to 10000 points and numerically checked that the our results are stable under perturbations of the input matrices $\epsilon\simeq {\cal O}(10^{-4})$. As a consequence we consider our results reliable as long as in Figs.~\ref{fig:relaxing_factor_20}--\ref{fig:relaxing_factor_1000} the horizontal bar representing $\xi^{1/2}$ does not exceed ${\cal O}(10^{2})$. Calculating WIMP--nucleus expected scattering rates with a larger precision would be extremely time consuming and probably difficult to achieve, given that they involve triple numerical integrals (on the recoil energy $E_R$, the visible energy $E^{\prime}$ and the WIMP incoming speed $v$, see Eqs.~(\ref{eq:start}, \ref{eq:start2}, \ref{eq:start3}, \ref{eq:start1})). As one can see from Figs.~\ref{fig:relaxing_factor_20}, \ref{fig:relaxing_factor_100} and \ref{fig:relaxing_factor_1000}, when interferences involving only contact or long range interactions are allowed the couplings for which $\xi^{1/2}\gsim {\cal O}(10^{2})$ (black horizontal bars) involve those driven by ${\cal O}_{11}$, ${\cal O}_{12}$ and ${\cal O}_{15}$, i.e.  $c^{n(p)}_{11}$, $c^{n(p)}_{12}$ and $c^{n(p)}_{15}$, $\alpha^{n(p)}_{11}$, $\alpha^{n(p)}_{12}$ and $\alpha^{n(p)}_{15}$. On the other hand, as shown by the horizontal red bars in the same figures, such excess of tuning affects also couplings driven by ${\cal O}_{1}$ and ${\cal O}_{3}$ when interferences in the full contact + long range parameter space of the Hamiltonian in Eq.~(\ref{eq:H_short_long}) are included, i.e. for $c^{n(p)}_{1}$, $c^{n(p)}_{3}$, $\alpha^{n(p)}_{1}$ and $\alpha^{n(p)}_{3}$. In such cases the numerical evaluation of the conservative bound should be taken with care, since it is likely to be unstable under numerical perturbations that go beyond the precision of our calculation.

\section{Conclusions}
\label{sec:conclusions}

It is customary to represent the null results of Weakly Interacting Massive Particles (WIMPs) direct detection searches with exclusion plots where the upper bound on the WIMP--nucleon spin--independent or spin--dependent cross section is provided as a function of the WIMP mass. In Refs.~\cite{conservative_icecube_2021,  complementarity_munich_sogang} a new method was introduced to calculate exclusion bands that bracket the exclusion plot for the couplings of the non--relativistic effective Hamiltonian for WIMP--nucleus scattering including the effect of interferences. Assuming a standard Maxwellian velocity distribution for the WIMPs in the halo of our Galaxy in the present paper we have applied such procedure to the null results of an exhaustive set of 9 direct detection experiments to calculate the exclusion bands for each of the Wilson coefficients of the effective Hamiltonian for a WIMP of spin 1/2. We have considered 56 Wilson coefficients $c_i^{p,n}$ and $\alpha_i^{n,p}$ for the WIMP--proton and WIMP--neutron contact interactions ${\cal O}_i^{p,n}$ and the corresponding long range interactions ${\cal O}_i^{p,n}/q^2$, parameterized by a massless propagator $1/q^2$. We provided a different exclusion band when each of the following set of operators was allowed to interfere: proton--neutron, i.e. $c_i^{p}$--$c_i^{n}$ or $\alpha_i^{p}$--$\alpha_i^{n}$; contact-contact or long range--long range, i.e. $c_i^{p,n}$--$c_j^{p,n}$ or $\alpha_i^{p,n}$--$\alpha_j^{p,n}$; contact--long range, i.e. $c_i^{p,n}$--$\alpha_j^{p,n}$. 
The procedure described above is complicated by the fact that some of the matrices that enter the calculation of WIMP--nucleus scattering in non--relativistic effective theory have flat directions and are close to singular~\cite{complementarity_munich_sogang}. This can lead to numerical instabilities that can be avoided by combining the constraints of different experiments. However for this to work it is crucial that the target nuclei are complementary, so that more than one experiment determines the bound (i.e. the corresponding ellipsoids intersect) and that the flat directions of the different targets do not overlap. In Appendix~\ref{app:factorization} we have provided some semi--analytical arguments to explain where flat directions are coming from and why they are expected to mostly affect experiments that use a single target and a reduced energy range.  

Our main quantitative results are shown in Figs.~\ref{fig:SI_shade}--\ref{fig:no_op_interf_long_shade}, where for each of the 56 Wilson coefficients $c_i^{p,n}$ and $\alpha_j^{p,n}$ the  exclusion band is plotted as a function of the WIMP mass $m_\chi$ when the corresponding operator is allowed to interfere with a growing number of other interactions.
For all the couplings we found that only three targets (xenon in LZ or PandaX--4T, fluorine in PICO--60($C_3F_8$) and PICO--60 ($CF_3I$), and iodine in PICO--60 ($CF_3I$)) contribute to determine $\max(|c_\alpha|)$.  

The width of the exclusion bands can reach 3 orders of magnitude and reduces to a factor as small as a few for the Wilson coefficients of the effective interactions where the WIMP couples to the nuclear spin, thanks to the complementarity between experiments that use proton--odd and neutron--odd targets. 

We have pointed out that some of the conservative bounds require an extremely high level of cancellation, which is at least as large as the square of the relaxation factor between the most constraining and the most conservative upper bound, but in some cases can be significantly larger. We have analyzed this issue in a systematic way in Figs.~\ref{fig:relaxing_factor_20}, \ref{fig:relaxing_factor_100} and \ref{fig:relaxing_factor_1000} showing that this problem affects some of the couplings driven by the operators  ${\cal O}_{1}$, ${\cal O}_{3}$, ${\cal O}_{11}$, ${\cal O}_{12}$ and ${\cal O}_{15}$, especially when interferences among contact and long range interactions are considered. For such couplings the reliability of the result is questionable, since the latter is very sensitive to extremely small changes in the input values of the matrices. On the other hand, it is possible to bracket the exclusion plot of the other couplings (${\cal O}_{4}$, ${\cal O}_{5}$, ${\cal O}_{6}$, ${\cal O}_{7}$, ${\cal O}_{8}$, ${\cal O}_{9}$, ${\cal O}_{10}$, ${\cal O}_{13}$, ${\cal O}_{14}$) to a narrow range in a robust way. This assessment is allowed by the fact that the semi–-analytic approach used in the present analysis allows to check in a straightforward way if the optimization procedure has converged, clearly setting apart numerically stable scenarios from those
that are unstable. 
%In particular in all the plots at low WIMP masses such convergence is systematically not achieved. For this reason in all of them we put a cut at $m_\chi$ = 15 GeV.
In this way we have also observed, in agreement with the conclusions of Ref.~\cite{complementarity_munich_sogang}, that at low WIMP masses the convergence of the optimization procedure is systematically not achieved. As a consequence we have removed configurations with $m_\chi\le$ 15 GeV from our plots.
We explain this with the fact that when $m_\chi$ is small all signals are suppressed by the tail of the velocity distribution and their sensitivity to the parameters is enhanced. This implies that the method discussed in the present paper is not suitable to obtain conservative bounds for light WIMPs. However in this mass range the signals are very sensitive to the details of the high--speed tail of the WIMP velocity distribution, which is affected by large uncertainties, so a combination of the present method with a halo--independent approach~\cite{Gondolo_Scopel_2017, halo_independent_Kahlhoefer, Halo-independent_Ferrer2015} would be probably more sensible to properly bracket the exclusion plot.

The only strategy to solve or alleviate the problem of large cancellations affecting the ${\cal O}_{1}$, ${\cal O}_{3}$, ${\cal O}_{11}$, ${\cal O}_{12}$ and ${\cal O}_{15}$ effective operators is to better exploit the complementarity of different targets. In particular our analysis has shown that for $m_\chi\ge$ 15 GeV only WIMP scattering off $Xe$, $F$ or $I$ plays a role in determining the allowed parameter space with the constraints of existing experiments. To improve existing bounds will require to increase the sensitivity of experiments that use other targets, or to add new nuclear targets for the use in direct detection.

\section*{Acknowledgements}
This research was supported by the National
Research Foundation of Korea(NRF) funded by the Ministry of Education
through the Center for Quantum Space Time (CQUeST) with grant number
2020R1A6A1A03047877 and by the Ministry of Science and ICT with grant
number 2021R1F1A1057119.

\appendix

\section{Scattering squared amplitude}
\label{app:factorization}

We provide here for completeness the WIMP response functions $R_k^{\tau\tau^{\prime}}$ adapted from \cite{nreft_haxton2}.

\begin{eqnarray}
 R_{M}^{\tau \tau^\prime}\left(v_T^{\perp 2}, {q^2 \over m_N^2}\right) &=& c_1^\tau c_1^{\tau^\prime } + {j_\chi (j_\chi+1) \over 3} \left[ {q^2 \over m_N^2} v_T^{\perp 2} c_5^\tau c_5^{\tau^\prime }+v_T^{\perp 2}c_8^\tau c_8^{\tau^\prime }
+ {q^2 \over m_N^2} c_{11}^\tau c_{11}^{\tau^\prime } \right] \nonumber \\
 R_{\Phi^{\prime \prime}}^{\tau \tau^\prime}\left(v_T^{\perp 2}, {q^2 \over m_N^2}\right) &=& \left [{q^2 \over 4 m_N^2} c_3^\tau c_3^{\tau^\prime } + {j_\chi (j_\chi+1) \over 12} \left( c_{12}^\tau-{q^2 \over m_N^2} c_{15}^\tau\right) \left( c_{12}^{\tau^\prime }-{q^2 \over m_N^2}c_{15}^{\tau^\prime} \right)\right ]\frac{q^2}{m_N^2}  \nonumber \\
 R_{\Phi^{\prime \prime} M}^{\tau \tau^\prime}\left(v_T^{\perp 2}, {q^2 \over m_N^2}\right) &=& \left [ c_3^\tau c_1^{\tau^\prime } + {j_\chi (j_\chi+1) \over 3} \left( c_{12}^\tau -{q^2 \over m_N^2} c_{15}^\tau \right) c_{11}^{\tau^\prime }\right ] \frac{q^2}{m_N^2} \nonumber \\
  R_{\tilde{\Phi}^\prime}^{\tau \tau^\prime}\left(v_T^{\perp 2}, {q^2 \over m_N^2}\right) &=&\left [{j_\chi (j_\chi+1) \over 12} \left ( c_{12}^\tau c_{12}^{\tau^\prime }+{q^2 \over m_N^2}  c_{13}^\tau c_{13}^{\tau^\prime}  \right )\right ]\frac{q^2}{m_N^2} \nonumber \\
   R_{\Sigma^{\prime \prime}}^{\tau \tau^\prime}\left(v_T^{\perp 2}, {q^2 \over m_N^2}\right)  &=&{q^2 \over 4 m_N^2} c_{10}^\tau  c_{10}^{\tau^\prime } +
  {j_\chi (j_\chi+1) \over 12} \left[ c_4^\tau c_4^{\tau^\prime} + \right.  \nonumber \\
 && \left. {q^2 \over m_N^2} ( c_4^\tau c_6^{\tau^\prime }+c_6^\tau c_4^{\tau^\prime })+
 {q^4 \over m_N^4} c_{6}^\tau c_{6}^{\tau^\prime } +v_T^{\perp 2} c_{12}^\tau c_{12}^{\tau^\prime }+{q^2 \over m_N^2} v_T^{\perp 2} c_{13}^\tau c_{13}^{\tau^\prime } \right] \nonumber \\
    R_{\Sigma^\prime}^{\tau \tau^\prime}\left(v_T^{\perp 2}, {q^2 \over m_N^2}\right)  &=&{1 \over 8} \left[ {q^2 \over  m_N^2}  v_T^{\perp 2} c_{3}^\tau  c_{3}^{\tau^\prime } + v_T^{\perp 2}  c_{7}^\tau  c_{7}^{\tau^\prime }  \right]
       + {j_\chi (j_\chi+1) \over 12} \left[ c_4^\tau c_4^{\tau^\prime} +  \right.\nonumber \\
       &&\left. {q^2 \over m_N^2} c_9^\tau c_9^{\tau^\prime }+{v_T^{\perp 2} \over 2} \left(c_{12}^\tau-{q^2 \over m_N^2}c_{15}^\tau \right) \left( c_{12}^{\tau^\prime }-{q^2 \over m_N^2}c_{15}^{\tau \prime} \right) +{q^2 \over 2 m_N^2} v_T^{\perp 2}  c_{14}^\tau c_{14}^{\tau^\prime } \right] \nonumber \\
     R_{\Delta}^{\tau \tau^\prime}\left(v_T^{\perp 2}, {q^2 \over m_N^2}\right)&=& {j_\chi (j_\chi+1) \over 3} \left( {q^2 \over m_N^2} c_{5}^\tau c_{5}^{\tau^\prime }+ c_{8}^\tau c_{8}^{\tau^\prime } \right)\frac{q^2}{m_N^2} \nonumber \\
 R_{\Delta \Sigma^\prime}^{\tau \tau^\prime}\left(v_T^{\perp 2}, {q^2 \over m_N^2}\right)&=& {j_\chi (j_\chi+1) \over 3} \left (c_{5}^\tau c_{4}^{\tau^\prime }-c_8^\tau c_9^{\tau^\prime} \right) \frac{q^2}{m_N^2}.
\label{eq:wimp_response_functions}
\end{eqnarray}

In the expressions above the Wilson coefficients $c^\tau_i$ can be generic functions of the transferred momentum $q$. In particular for the Hamiltonian of Eq.~(\ref{eq:H_short_long}) the contact interaction part 
is obtained by taking constant $c^\tau_i$'s, while the long range interaction contribution is obtained with the substitution $c^\tau_i\rightarrow \alpha^\tau_i/q^2$.

\subsection{Singularity of squared amplitude matrices} 
\label{sec:sensitivity}

In this Section we wish to provide some arguments to clarify the origin of the fact that some of the matrices ${\cal M}$ for the calculation of WIMP--nucleus scattering are close to singular. 

The calculation of the squared amplitude of Eq.~(\ref{eq:squared_amplitude}) within the non--relativistic effective theory of the nucleon scattering of a WIMP of spin 1/2 was originally provided in Refs.~\cite{nreft_haxton1,nreft_haxton2} (more details can be found in~\cite{all_spins_theory}, where such derivation was extended to WIMPs of arbitrary spin). In both cases the evaluation is in one--nucleon approximation, i.e. the WIMP is assumed to interact with a single nucleon at a time inside the nucleus. Besides allowing a nice factorization between the physics of the nucleus and that of the WIMP particle, such approximation implies that for a given nuclear target $T$ the functions $W_{T,XY}^{\tau\tau^{\prime}}(q)$ (indicated with $W_{T,X}^{\tau\tau^{\prime}}(q)$ when $X$ = $Y$ and where $XY$ = $M$, $\Phi^{\prime\prime}$, $\Phi^{\prime\prime}M$, $\tilde{\Phi}^{\prime}$,
$\Sigma^{\prime\prime}$, $\Sigma^{\prime}$, $\Delta$, $\Delta\Sigma^{\prime}$) can be factorized as:

\begin{equation}
    W_{XY}^{\tau\tau^{\prime}}(q)=W_X^{\tau}(q) W_Y^{\tau^{\prime}}(q),
    \label{eq:W_factorization}
\end{equation}

\noindent and are singular two--by--two dimensional matrices in isospin space. In particular in Refs.~\cite{nreft_haxton2, Catena_nuclear_form_factors} numerical approximations of the $W^{\tau\tau^{\prime}}_k$ form factors are provided for each combination $\tau$, $\tau^{\prime}$. We notice here that in light of Eq.~(\ref{eq:W_factorization}) such numerical evaluations are redundant, in the sense that only the 2 components of $W_X^\tau$ out of the 4 components of $W^{\tau\tau^{\prime}}_X$ are independent. By the same token, out of a total of 12 components $W_X^{\tau\tau^\prime}$, $W_Y^{\tau\tau^\prime}$ and $W_{XY}^{\tau\tau^\prime}$ for two interfering nuclear form factors $X$ and $Y$ only the 4 four components of $W_X^\tau$, $W_Y^\tau$ are independent. For instance, in the latter case, the 12 form factors contained in the 3 corresponding matrices can be written in terms of the four functions $W_X^{00}$, $W_X^{01}$, $W_{XY}^{00}$ and $W_{XY}^{01}$:

\begin{eqnarray}
&&W^{\tau\tau^{\prime}}_X(q)=\left ( \begin{array}{cc} W_X^{00} & W_X^{01}\\ W_X^{01} & \frac{(W_X^{01})^2}{W_X^{00}} \end{array}\right ),\,\,\,
W^{\tau\tau^{\prime}}_Y(q)=\left ( \begin{array}{cc} \frac{(W_{XY}^{00})^2}{W_X^{00}} & \frac{W_{XY}^{00}W_{XY}^{01}}{W_X^{00}}\\ \frac{W_{XY}^{00}W_{XY}^{01}}{W_X^{00}} & \frac{(W_{XY}^{01})^2}{W_X^{00}} \end{array}\right ),\nonumber\\
&& W^{\tau\tau^{\prime}}_{XY}(q)=\left ( \begin{array}{cc} W_{XY}^{00} & W_{XY}^{01}\\ \frac{W_X^{01}W_{XY}^{00}}{W_X^{00}} & \frac{W_X^{01} W_{XY}^{01}}{W_X^{00}} \end{array}\right ),
    \label{eq:w_parameterization}
    \end{eqnarray}

\noindent with:
\begin{equation}
    W^0_X=\sqrt{W^{00}_X};\,\,\,W^1_X=\frac{W_X^{01}}{\sqrt{W_X^{00}}},\,\,\,W^0_Y=\frac{W^{00}_{XY}}{\sqrt{W^{00}_X}};\,\,\,W^1_Y=\frac{W_{XY}^{01}}{\sqrt{W_X^{00}}}.
\end{equation}

When the factorization (\ref{eq:W_factorization}) is substituted explicitly in the expression of the squared amplitude of Eq.~(\ref{eq:squared_amplitude}) using (\ref{eq:wimp_response_functions}) one gets (assuming for definiteness a contact interaction):

\begin{equation}
    \frac{1}{2 j_{\chi}+1} \frac{1}{2 j_{T}+1}|\mathcal{M}|^2=\frac{4\pi}{2j_T+1}\left [A^{(0)}+A^{(1)}(v^2-v_{min}^2)\right ]=\bm{c}^T\cdot {\cal A}\cdot \bm{c},
\label{eq:amplitude_a0_a2}
\end{equation}

\noindent where the quantities $A^{(0)}$ and $A^{(1)}$ are given by:

\begin{eqnarray}
 A^{(0)}&=& A_{[1,3]}^2+A_{[10]}^2+\frac{j_\chi(j_\chi +1)}{12}\left(A_{[4,6]}^2+A_{[4,5]}^2+A_{[8,9]}^2+A_{[11,12,15]}^2+(A_{[12]}^{(0)})^2+(A_{[13]}^{(0)})^2 \right)  \nonumber\\
 A^{(1)}&=&A_{[3]}^2+A_{[7]}^2+\frac{j_\chi(j_\chi +1)}{3}\left (A_{[5]}^2+A_{[8]}^2+(A_{[12]}^{(1)})^2+A_{[12,15]}^2+(A_{[13]}^{(1)})^2+A_{[14]}^2 \right )
 \label{eq:amplitude_polynomials}
\end{eqnarray}
\noindent In the expression above the $A_{[n_1,...,n_k]}$ quantities are polynomials linear in the couplings $c_{n_1}^\tau$,...$c_{n_k}^\tau$:

\begin{eqnarray}
 A_{[1,3]}&=& c_1 \cdot W_M+\frac{1}{2}\tilde{q}^2 c_3\cdot W_{\Phi^{\prime\prime}}  \nonumber\\
 A_{[10]}&=& \frac{1}{2}\tilde{q} c_{10}\cdot W_{\Sigma^{\prime\prime}}  \nonumber\\
 A_{[4,6]}&=& c_4\cdot W_{\Sigma^{\prime\prime}}+\tilde{q}^2c_6\cdot W_{\Sigma^{\prime\prime}}  \nonumber\\
 A_{[4,5]}&=& c_4\cdot W_{\Sigma^{\prime}}+2\tilde{q}^2 c_5\cdot W_\Delta \nonumber\\
 A_{[8,9]}&=& 2\tilde{q} c_8\cdot W_\Delta-\tilde{q}c_9\cdot W_{\Sigma^{\prime}} \nonumber\\
 A_{[11,12,15]}&=& 2\tilde{q} c_{11}\cdot W_M+\tilde{q}(c_{12}-\tilde{q}^2 c_{15})\cdot W_{\Phi^{\prime\prime}}  \nonumber\\
  A_{[12]}^{(0)}&=& \tilde{q}c_{12}\cdot W_{\tilde{\Phi}^{\prime}} \nonumber\\
  A_{[13]}^{(0)}&=& \tilde{q}^2 c_{13}\cdot W_{\tilde{\Phi}^{\prime}}  \nonumber\\
  A_{[3]}&=& \frac{1}{\sqrt{8}}\tilde{q} c_3\cdot W_{\Sigma^{\prime}}\nonumber\\
  A_{[7]}&=& \frac{1}{\sqrt{8}}c_7\cdot W_{\Sigma^{\prime}}\nonumber\\
  A_{[5]}&=& \tilde{q} c_5\cdot W_M\nonumber\\
 A_{[8]}&=& c_8\cdot W_M\nonumber\\
 A_{[12]}^{(1)}&=& \frac{1}{2}  c_{12}W_{\Sigma^{\prime\prime}}\nonumber\\
  A_{[12,15]}&=& \frac{1}{\sqrt{8}} (c_{12}-\tilde{q}^2 c_{15})\cdot W_{\Sigma^{\prime}}\nonumber\\
  A_{[13]}^{(1)}&=& \frac{1}{2} \tilde{q} c_{13}\cdot W_{\Sigma^{\prime\prime}}\nonumber\\
  A_{[14]}&=& \frac{1}{2}\tilde{q} c_{14} W_{\Sigma^{\prime}} 
  \nonumber\label{eq:polynomials}\\
\end{eqnarray}
\noindent where $\tilde{q}\equiv q/m_N$ and $c_n\cdot W_X$ = $c_n^0 W_X^0+c_n^1 W_X^1$.
In the equations above the Wilson coefficients can be arbitrary functions of $q^2$, so with the substitution $c_i^\tau\rightarrow c_i^\tau+\alpha_i^\tau/q^2$ they can be directly used in the full 56--dimensional parameter space of the Hamiltonian~(\ref{eq:H_short_long}).

Eq.~(\ref{eq:amplitude_polynomials}) shows that  all the non--interfering blocks of the matrix ${\cal A}$ in Eq.~(\ref{eq:amplitude_a0_a2}) are singular. In fact in any couplings subspace both $A^{(0)}$ and $A^{(1)}$ are set to zero by a number of linear conditions (one for each polynomial in the sum (\ref{eq:amplitude_polynomials})) that is smaller than the dimensionality of the subspace, so that the kernel of the ${\cal A}$ submatrix has dimension larger than zero.  For instance, considering only the contact interactions in the velocity--independent part $A^{(0)}$ of the squared amplitude the subspace $[4,5,6]$ has dimensionality 6, but in Eq.~(\ref{eq:amplitude_polynomials}) only two polynomials depend on the corresponding couplings. In this subspace the 4--dimensional kernel of ${\cal A}$ is spanned by all the coupling vectors $\bm{c}$ = $[c_4^0, c_4^1, c_5^0, c_5^1,c_6^0, c_6^1, ]$ perpendicular to $\bm{c}_1$ = $[W_{\Sigma^{\prime\prime}}^0,W_{\Sigma^{\prime\prime}}^1, 0,0, \tilde{q}^2 W_{\Sigma^{\prime\prime}}^0, W_{\Sigma^{\prime\prime}}^1]$ and $\bm{c}_2$ = $[W_{\Sigma^{\prime}}^0,W_{\Sigma^{\prime}}^1, 2\tilde{q}^2 W_\Delta^0, W_\Delta^1, 0, 0]$. By inspection it is possible to verify that the same happens in all the non--interfering subspaces of the matrix ${\cal A}$.

The discussion above implies that, since $(dR/dE^{\prime})_T\propto {\cal A}$, in the effective theory parameter space the differential rate at fixed visible energy $E^{\prime}$ on a single target $T$ (see Eq.~(\ref{eq:diff_rate_eprime})) is not only given by a singular matrix, but its kernel can have a large dimensionality. In other words, the quadratic form ${\cal R}_{diff}$ of the differential rate has always flat directions extending to infinity. As a consequence, if the differential rate at fixed energy and on a single target were to be used to put constraints on the model, the condition $(dR/dE^{\prime})_T$ = $\bm{c}^t\cdot {\cal R}_{diff} \cdot \bm{c}<(dR/dE^{\prime})_{max}$  would yield a divergent conservative bound on  $c_\alpha$ because ${\cal R}_{diff}$ is singular.

In a realistic set-up the differential rate is always integrated in some energy range $E^{\prime}_1<E^{\prime}<E^{\prime}_2$ and in some cases the total rate is given by the sum of contributions off different targets $T$ (either different isotopes of the same species, or different nuclei in the case of molecular targets). In both cases
provided that the matrix ${\cal R}(E^{\prime})$ varies enough in the $[E^{\prime}_1,E^{\prime}_2]$ interval and/or the targets $T$ have different flat directions the matrix ${\cal R}_{Exp\, 1}$ in Fig.~\ref{fig:ellipses_bracketing} is not singular.
In this case the matrix eigenvalues $\lambda_i$ do not vanish anymore but a huge numerical hierarchy can survive among them if a single target contributes to the expected rate, and when the latter takes contribution from a reduced range of energies.  In such case the semi--major axes of the corresponding ellipsoids are proportional to $\lambda_i^{-1/2}$ and in practice the exact extension of the ellipsoid along the (almost) flat direction for which $\lambda_i\rightarrow$ 0 becomes extremely sensitive to small corrections of the matrix entries. We observe this effect in the case of the PICO--60 experiments using $C_3F_8$:  for interactions of the "spin--dependent" type scattering off $^{12}C$ vanishes and the rate is only driven by WIMP scatterings off $^{19}F$, while the range of recoil energies contributing to the expected rate is relatively small due to the low mass of the target.

\section{Implementations of experiments}
\label{app:experiments}
%%%%%%%%%%%%%%%%%%%%%%%%%%%%%%%%%%%%%%%%%%%%%
\subsection{LZ, PandaX--4T and XENON1T}
\label{app:lz}

For LUX-ZEPLIN (LZ) we assume an exposure of 3.3$\times$10$^5$ kg days, the nuclear recoil energy range 1.25 keV $\le E_R \le$ 80 keV~\cite{LZ_2022} and the efficiency provided in Fig.~2 of~\cite{LZ_2022}. For PandaX--4T we assume an exposure 0.63 tonne year, the recoil energy range 2 keV $\le E_R \le$ to 135 keV~\cite{pandax4T_2021} and the efficiency provided in Fig.~2 of~\cite{pandax4T_2021}. We reproduce the published exclusion plots reasonably well for a standard SI interaction assuming 3.4 and 3.6 residual candidate events for LZ and PandaX--4T, respectively, and we use such values to obtain all our exclusion plots.

For XENON1T we assume an exposure of 3.6$\times$10$^{5}$ kg days, 7 events in the nuclear recoil energy range 1.8 keV $\le E_R \le$ 62 keV~\cite{xenon_2018} and the efficiency provided in Fig.~1 of~\cite{xenon_2018}. For all three experiments the provided efficiencies are directly expressed in keV and include the effects of quenching and energy resolution.

%%%%%%%%%%%%%%%%%%%%%%%%%%%%%%%%%%%%%%%%%%%%%

%\subsection{XENON1T}
%\label{app:xenon1t}
%The published XENON1T limit of of Ref.~\cite{xenon_2018} is the result of a non--trivial statistical analysis including many nuisance parameters that cannot be accurately reproduced by simply comparing the measured experimental upper bound on the number of WIMP--candidate events in some energy bins.
%Assuming 7 WIMP candidate events in the range of PE 1.8 $ \le S_1 \le $ 62, we could reproduce relatively well the published bound, as shown in Fig.~3 of Ref.~\cite{xenon_2018}
%for the primary scintillation signal S1 (directly in Photo Electrons,
%PE), with an exposure of 278.8 days and a fiducial volume of 1.3 ton
%of xenon. This allows us to directly convolute the efficiency provided in Fig.~1
%of~\cite{xenon_2018}, which includes the effects of quenching and energy resolution, 
%with the differential rate of Eq.~(\ref{eq:dr_de}) to obtain the expected rate.
%changed eq:dr_der to eq:dr_de

%%%%%%%%%%%%%%%%%%%%%%%%%%%%%%%%%%%%%%%%%%%%%

\subsection{PICO--60 ($C_3F_8$)}
\label{app:pico60_c3f8}
Bubble chambers are threshold experiments that detect a signal only 
above some value $E_{th}$ of the deposited
energy. In this case the expected number of
events is given by:

\begin{equation}
R=N_T MT\int_0^{\infty} P(E_R) \frac{dR}{dE_R} dE_R,
\label{eq:r_threshold}
\end{equation}

\noindent with $P(E_R)$ the nucleation probability.

One of the target materials used by PICO--60 is $C_3F_8$, for which we
used the complete exposure~\cite{pico60_2019} consisting in 1404 kg
day at threshold $E_{th}$=2.45 (with 3 observed candidate events and 1
event from the expected background, implying an upper bound of 6.42
events at 90\%C.L.~\cite{feldman_cousin}) and 1167 kg day keV at
threshold $E_{th}$=3.3 keV (with zero observed candidate events and
negligible expected background, implying a 90\% C.L. upper bound of
2.3 events). For the two runs we have assumed the nucleation
probabilities in Fig. 3 of \cite{pico60_2019}. 
%%%%%%%%%%%%%%%%%%%%%%%%%%%%%%%%%%%%%%%%%%%%%

\subsection{PICO--60 ($CF_3I$)}
\label{app:pico60cf3i}

PICO--60 can also employ a $CF_3I$ target. For the analysis of
Ref.\cite{pico60_2015} we adopt an energy threshold of 13.6 keV and an
exposure of 1335 kg days. The nucleation probabilities for each target
element are taken from Fig.4 in~\cite{pico60_2015}.
%%%%%%%%%%%%%%%%%%%%%%%%%%%%%%%%%%%%%%%%%%%%%

\subsection{SuperCDMS}
\label{app:supercdms}

The latest SuperCDMS analysis \cite{super_cdms_2017} observed 1 event
between 4 and 100 keVnr with an exposure of 1690 kg days. We have
taken the efficiency from Fig.1 of \cite{super_cdms_2017} and the
energy resolution $\sigma=\sqrt{0.293^2+0.056^2 E_{ee}}$ from
\cite{cdms_resolution}.
%%%%%%%%%%%%%%%%%%%%%%%%%%%%%%%%%%%%%%%%%%%%%

\subsection{CDMSlite}
\label{app:cdmslite}

For CDMSlite we considered the energy bin of 0.056 keV$<E^{\prime}<$
1.1 keV with a measured count rate of 1.1$\pm$0.2 [keV kg day]$^{-1}$
(Full Run 2 rate, Table II of Ref. \cite{cdmslite_2017}). We have
taken the efficiency from Fig.4 of \cite{cdmslite_2017} and the energy
resolution $\sigma=\sqrt{\sigma_E^2+B E_R+(A E_R)^2}$, with
$\sigma_E$=9.26 eV, $A$=5.68$\times 10^{-3}$ and $B$=0.64 eV from
Section IV.A of~\cite{cdmslite_2017}.

\subsection{COSINE--100}
\label{app:cosine}

The exclusion plot for COSINE--100~\cite{cosine_nature} relies on a
Montecarlo~\cite{cosine_bck} to subtract the different backgrounds of
each of the eight crystals used in the analysis. In
Ref.~\cite{cosine_nature} the amount of residual background after
subtraction is not provided, so we have assumed a constant background
$b$ at low energy (2 keVee$< E_{ee}<$ 8 keVee), and estimated $b$ by
tuning it to reproduce the exclusion plot in Fig.4 of
Ref.~\cite{cosine_nature} for the isoscalar spin-independent elastic
case. The result of our procedure yields $b\simeq$0.13
events/kg/day/keVee, which implies a subtraction of about 95\% of the
background.  We take the energy resolution $\sigma/\mbox{keV}=0.3171
\sqrt{E_{ee}/\mbox{keVee}}+0.008189 E_{ee}/\mbox{keVee}$ averaged over
the COSINE--100 crystals~\cite{cosine_private} and the efficiency for
nuclear recoils from Fig.1 of Ref.~\cite{cosine_nature}. Quenching
factors for sodium and iodine are assumed to be equal to 0.3 and 0.09
respectively.

%%%%%%%%%%%%%%%%%%%%%%%%%%%%%%%%%%%%%%%%%%%%%
%\subsection{PandaX--4T}
%\label{app:pandax}
%
%PandaX--4T is a liquid xenon chamber with exposure 0.63 tonne year. We assume zero WIMP candidate events in the recoil energy range 2 keV $\le E_R \le$ to 135 keV~\cite{pandax4T_2021} and the efficiency provided in Fig.~2 of~\cite{pandax4T_2021}, which is directly in keV and includes the effects of quenching and energy resolution.
%
%
%%%%%%%%%%%%%%%%%%%%%%%%%%%%%%%%%%%%%%%%%%%%%
\subsection{DAMIC}
\label{app:damic}

The dark matter in CCDs~\cite{DAMIC_2016} experiment (DAMIC) employs a silicon target.
For our analysis we used the CCD 1$\times$100 spectrum from Fig.~10 of~\cite{DAMIC_2016} with exposure 0.204 kg days.
We take the quenching factor from Fig.~11 of~\cite{Izraelevitch2017} with a cut below 0.3keV, the efficiency from Fig.~9 of~\cite{DAMIC_2016} and the energy resolution $\sigma^2= \sigma^2_0 + (3.77 \mbox{eVee})~FE_{ee}$, with $F=0.133$ and $\sigma_0=30$ eVee.

%%%%%%%%%%%%%%%%%%%%%%%%%%%%%%%%%%%%%%%%%%%%%
\section{The wimp\_dd\_matrix routine in WimPyDD} 
\label{app:wimpydd}

All the calculations of the present paper have been performed using the \verb|WimPyDD|~\cite{wimpydd_2022} code, available at \href{https://wimpydd.hepforge.org/}{https://wimpydd.hepforge.org}.
In particular the matrices were produced using the \verb|wimp_dd_matrix| routine, that has been released in a new version of the code, in correspondence to this publication. For this reason we provide a short introduction to it in this Appendix. 

The routine \verb|wimp_dd_matrix| takes as input the WIMP mass \verb|m_chi|, an \verb|experiment| object \verb|exp| containing all the information of the experimental set--up (target, energetic bins, energy resolution, efficiency, exposure, etc.), a \verb|hamiltonian| object belonging to the \verb|eft_hamiltonian| class and the halo function $\eta(v_{min})=\int_{v_{min}}^\infty f(v)/v dv = \sum_{k=1}^{N_s}
  \delta\eta_k(t)\Theta(v_k-v_{min})$ stored in the two arrays \verb|vmin|, \verb|delta_eta|.    

For instance, in order to calculate the 8$\times$8--dimensional matrix for the subspace ($c^0_1$, $c^1_1$, $\alpha^0_1/q^2$, $\alpha^1_1/q^2$, $c^0_3$, $c^1_3$, $\alpha^0_3/q^2$, $\alpha^1_3/q^2$) generated by the interfering operators ${\cal O}_1$ and ${\cal O}_3$ and for $m_\chi$= 100 GeV one needs to input the following instructions:

\begin{Verbatim}[frame=single,xleftmargin=1cm,xrightmargin=1cm,commandchars=\\\{\}]
import WimPyDD as WD\\
import numpy as np\\
wc=\{1: lambda: [1,1], 3: lambda: [1,1], \\
(1,'qm2'): lambda q : [1/q**2,1/q**2],\\
(3,'qm2'): lambda q : [1/q**2,1/q**2]\}
hamiltonian=WD.eft_hamiltonian('model_1_3', wc)
n_bin=0
mchi=100
vmin,delta_eta=WD.streamed_halo_function()
m=WD.wimp_dd_matrix(exp, hamiltonian, n_bin, vmin, \\
delta_eta, mchi)
rotation=WD.rotation_from_isospin_to_pn(hamiltonian)\\
m=np.dot(rotation,np.dot(m,rotation))
mapping=WD.get_mapping(hamiltonian, pn=True)
mapping[3,'n']
3
mapping[(1,'qm2'),'p']
4
\end{Verbatim}

\noindent To obtain the \verb|experiment| object for XENON1T one can use  \verb|WD.XENON1T|, which is built--in in \verb|WimPyDD| and is implemented as described in~\ref{app:lz}. To calculate the halo function the routine \verb|WD.streamed_halo_function| is used  (passing no input arguments corresponds to a standard isotropic Maxwellian with default parameters). 

As explained in Section 3.4 of~\cite{wimpydd_2022} \verb|WimPyDD| handles the response functions generated by same interaction operators with different momentum dependences by extending the couplings keys with arbitrary string identifiers. For instance, in the example above the \verb|wc| dictionary keys \verb|1| and \verb|(1,'qm2')| correspond to $c_1^\tau$ and $\alpha^\tau_1/q^2$. 
The values of the \verb|wc| dictionary are arbitrary functions that return a two--dimensional array with the isospin components ($\tau$ =0 and 1) of a Wilson coefficient in GeV$^{-2}$.
To factor out the couplings from the matrix elements one can set the dictionary values to \verb|[1,1]| and \verb|[1/q**2,1/q**2]|, although any other normalization is possible. The parameter \verb|n_bin| selects one of the energy bins contained in the \verb|data.tab| file that initializes \verb|experiment| (see Ref.~\cite{wimpydd_2022} for details). Since \verb|WD.XENON1T| is implemented with a single energy bin \verb|n_bin=0|. The routine \verb|wimp_dd_matrix| takes as default $j_\chi$=1/2, but allows to take an arbitrary spin of the WIMP using the base of~\cite{all_spins_theory} in the \verb|wc| dictionary.

The matrix \verb|m| is calculated in isospin base and can be rotated to the proton--neutron base using the array produced by \verb|WD.rotation_from_isospin_to_pn|. The dictionary \verb|mapping| contains the mapping between the couplings in the keys of \verb|wc| and the indices of \verb|m|. It is obtained in the proton--neutron base by setting \verb|pn=True| in the \verb|WD.get_mapping| routine (\verb|pn=False| by default). For instance \verb|mapping[3,'p']| $\rightarrow$ 3 yields the index corresponding to $c_3^p$ while \verb|mapping[(1,'qm2'),'n']| $\rightarrow$ 4 the index for $\alpha^n_1$, so that \verb|m[3,4]| corresponds to the $c_3^p$--$\alpha^n_1$ component of \verb|m|.

\end{document}